\documentclass[fleqn, usenatbib]{mnras}
\usepackage{newtxtext,newtxmath}

\usepackage[T1]{fontenc}
\usepackage{ae,aecompl}

\usepackage{amssymb}
\usepackage{amsmath}
\usepackage{mathtools}
\usepackage{appendix}

\usepackage[usenames, dvipsnames]{color}
\usepackage[normalem]{ulem}

\title[Clustered SNe in 3D, MHD]{The momentum budget of clustered supernova feedback in a 3D, magnetized medium}

\author[E. S. Gentry et al.]{
Eric S. Gentry,\textsuperscript{1}\thanks{E-mail: egentry@ucsc.edu}
Mark R. Krumholz,\textsuperscript{2}
Piero Madau,\textsuperscript{1, 3}
Alessandro Lupi\textsuperscript{4}
\\
\textsuperscript{1}Department of Astronomy and Astrophysics, University of California at Santa Cruz, 1156 High St, Santa Cruz, CA 95064, USA\\
\textsuperscript{2}Research School of Astronomy \& Astrophysics, Australian National University, Canberra, ACT 2611, Australia\\
\textsuperscript{3}Sorbonne Universit\'e, CNRS, UMR 7095, Institut d'Astrophysique de Paris, 98 bis bd Arago, F-75014 Paris, France\\
\textsuperscript{4}Scuola Normale Superiore, Piazza dei Cavalieri 7, I-56126 Pisa, Italy
}

\date{Accepted 2018 November 30. Received 2018 November 30; in original form 2018 February 6}

\pubyear{2019}

\begin{document}
\label{firstpage}
\pagerange{\pageref{firstpage}--\pageref{lastpage}}
\maketitle

\begin{abstract} 
While the evolution of superbubbles driven by clustered supernovae has been studied by numerous authors, the resulting radial momentum yield is uncertain by as much as an order of magnitude depending on the computational methods and assumed properties of the surrounding interstellar medium (ISM).
In this work, we study the origin of these discrepancies, and seek to determine the correct momentum budget for a homogeneous ISM. We carry out 3D hydrodynamic (HD) and magnetohydrodynamic (MHD) simulations of clustered supernova explosions, using a Lagrangian method and checking for convergence with respect to resolution. We find that the terminal momentum of a shell driven by clustered supernovae is dictated primarily by the mixing rate across the contact discontinuity between the hot and cold phases, and that this energy mixing rate is dominated by numerical diffusion even at the highest resolution we can complete, 0.03 $M_\odot$. Magnetic fields also reduce the mixing rate, so that MHD simulations produce higher momentum yields than HD ones at equal resolution. As a result, we obtain only a lower limit on the momentum yield from clustered supernovae. Combining this with our previous 1D results, which provide an upper limit because they allow almost no mixing across the contact discontinuity, we conclude that the momentum yield per supernova from clustered supernovae in a homogeneous ISM is bounded between $2\times 10^5$ and $3\times 10^6$ $M_\odot$ km s$^{-1}$.  A converged value for the simple homogeneous ISM remains elusive.

\end{abstract}



\section{Introduction}
\label{section:intro}

Feedback from supernovae (SNe) is an important component of understanding the interstellar medium (ISM), galactic winds, and galactic evolution \citep[e.g.,][]{1977ApJ...218..148M,1986ApJ...303...39D,2005ApJ...618..569M,2011ApJ...731...41O,2011ApJ...743...25K,2011ApJ...734...65J,2012MNRAS.421.3522H,2013MNRAS.432..455D,2013MNRAS.433.1970F,2013MNRAS.429.1922C,2016MNRAS.455..334T}. Unfortunately, the processes governing the strength of SN feedback operate non-linearly and at small scales. This makes it difficult to include the effects of SNe in analytic models or large galactic simulations without a simplified prescription for SN feedback.

In the past, most investigations of the key factors governing SN feedback strength have focused on single, isolated SNe \citep[e.g.,][]{1974ApJ...188..501C,1988ApJ...334..252C,1998ApJ...500...95T,2015A&A...576A..95I}.   In reality, however, core collapse SNe are clustered in space and time: massive stars are born in clusters, and explode after $\sim3-40$ Myr, before these stars can significantly disperse. The few studies that have looked at the feedback from multiple clustered, interacting SNe have found conflicting results. While some studies have found relatively small changes in the momentum from clustering SNe \citep{2015ApJ...802...99K,2015MNRAS.451.2757W, 2017ApJ...834...25K}, others have found that it could increase the average momentum per SN to 5-10 times greater than the traditional yields for isolated SNe \citep{2014MNRAS.442.3013K,2017MNRAS.465.2471G}.

It has been suggested that the differences in results for clustered SN simulations could stem from the different levels of mixing in the simulations, from both physical and non-physical sources.  Unfortunately, each recent simulation was idealized in significantly different ways, which makes it difficult for us to directly isolate which aspects were the primary drivers of the differences.  Our goal in this paper is to identify the causes of the discrepancies between different published results, and resolve whether, when including appropriate physics, clustering does in fact lead to significant changes in the terminal momentum of supernova remnants.

One of the key issues that we investigate is dimensionality and resolution. We found that clustering produces an order of magnitude enhancement in momentum  \citep{2017MNRAS.465.2471G}, but these results were based on 1D spherically symmetric simulations. Assuming spherical symmetry is potentially misleading because we know of fluid instabilities (such as the Vishniac instabilities and the Rayleigh--Taylor instability) that affect SNR morphologies \citep{1983ApJ...274..152V,1994ApJ...428..186V,1988ApJ...324..776M,1993ApJ...407..207M,2013A&A...550A..49K,2016MNRAS.456..710F}. Even small perturbations can be amplified and noticeably change key properties of SNR evolution.  For isolated SN simulations, 1D and 3D simulations do not produce significantly different terminal momenta (e.g., \citealt{2015MNRAS.450..504M}, \citealt{2015ApJ...802...99K}, and \citealt{2015MNRAS.451.2757W} all find differences of less than 60\% between 1D and 3D), but it is worth re-investigating the issue for clustered SNe. It could be that the longer time frame allows the instabilities to grow to have larger effects.

Conversely, our 1D simulations achieved much higher resolution than in any of the 3D simulations found in the literature. We found that the terminal momentum for clustered SNe did not converge until we reached peak resolutions better than $0.1$ pc, far higher than the resolutions of published 3D simulations. Moreover, we achieved this convergence only by using pseudo-Lagrangian methods that minimized numerical diffusion across the contact discontinuity at the inner edge of the superbubble, whereas many of the published 3D results are based on Eulerian methods that, for fronts advecting across the grid at high speed, are much more diffusive. Indeed, it is noteworthy that the one published 3D result that finds a significant momentum enhancement from clustering \citep{2014MNRAS.442.3013K} uses a Lagrangian method, while all the papers reporting no enhancement are based on Eulerian methods. Clearly, given the conjoined issues of resolution and dimensionality, further investigation is warranted.

Since the suppression and enhancement of mixing is a key unknown for the feedback budget of clustered SNe, we also explore the role of magnetic fields, which may reduce the amount of physical mixing.  Our interest in this possibility comes primarily from the example of cold fronts in galaxy clusters, where magnetic fields draped across a shock front have been used to explain the stability of these cold fronts against fluid instabilities (\citealt{2001ApJ...549L..47V,Markevitch07a}; although see also \citealt{2004MNRAS.350L..52C} who show that magnetic fields might not be necessary for stabilizing cold fronts).

In this paper, we first test the effects of bringing our simulations from 1D to 3D and carry out a 3D convergence study, and then we test the effects of adding magnetic fields into our 3D simulations. In \autoref{section:code}, we discuss our computational methods. In \autoref{section:results}, we discuss the results of our simulations, with a more detailed physical analysis of the significance of those results in \autoref{section:analysis}. In \autoref{section:conclusions}, we discuss our conclusions and compare to other works.


\section{Computational Methods}
\label{section:code}

For this work we repeat one of the 1D simulations from \citet{2017MNRAS.465.2471G}, and conduct 3D simulations of the same set-up at a range of resolutions and including or excluding magnetic fields. For the most part our 1D simulations reuse the code developed by \citet{2017MNRAS.465.2471G}, with minor changes that we discuss in \autoref{section:code:1D}.  In \autoref{section:code:3D}, we discuss the methodology for our 3D simulations, for which we use the \texttt{GIZMO} code \citep{2015MNRAS.450...53H,2016MNRAS.455...51H}.  We use \texttt{GIZMO} for this work because it has a Lagrangian hydrodynamic solver; in our previous 1D simulations, we found that Lagrangian methods were more likely to converge for simulations of clustered SNe \citep{2017MNRAS.465.2471G}.  

\subsection{1D simulation}
\label{section:code:1D}

The methods for our 1D simulation are very similar to those used in our previous work \citep{2017MNRAS.465.2471G}, with only slight modifications. First, we disable the injection of pre-SN winds, because injecting small amounts of mass over extended periods is impractical at the resolutions we are able to achieve in the 3D simulations. Second, we initialize the ISM to be at an equilibrium temperature ($T \sim 340$ K or a specific internal energy of $e_\mathrm{int} \sim 3.5 \times 10^{10}$ erg g$^{-1}$ for an initial ISM density of $\rho_0 = $ 1.33 $m_\mathrm{H}$ cm$^{-3}$ and gas-phase metallicity of $Z=0.02$, rather than $T \sim 15~000$ K).\footnote{Throughout this paper we will quote temperatures calculated by \texttt{GRACKLE} which accounts for temperature dependence in the mean molecular weight, $\mu$ \citep{2017MNRAS.466.2217S}.} This simplifies the analysis, as changes in energy now only occur in feedback-affected gas. Furthermore the initial temperature makes little difference as the gas would otherwise rapidly cool to its equilibrium state (the 15~000 K gas had a cooling time of a few kyr).  Using these modified methods we reran the most-studied cluster from our previous work, one that had a stellar mass of $M_\star = 10^3$ $M_\odot$ (producing 11 SNe) and was embedded in an ISM of initial density $\rho_0 = $ 1.33 $m_\mathrm{H}$ cm$^{-3}$ and an initial gas-phase metallicity of $Z=0.02$.\footnote{This cluster can be found in the tables produced by \citet{2017MNRAS.465.2471G} using the id \texttt{25451948-485f-46fe-b87b-f4329d03b203}.}  These changes allowed for more direct comparison with our 3D simulations, and do not affect our final conclusions.

The remainder of our methodology is identical to that of \citet{2017MNRAS.465.2471G}, which we summarize here for convenience. To generate a star cluster of given mass, we used the \texttt{SLUG} code \citep{2012ApJ...745..145D, 2014MNRAS.444.3275D, 2015MNRAS.452.1447K} to realistically sample a \citet{Kroupa04012002} IMF of stars. We assume every star with an initial mass above $8 M_\odot$ explodes as a core collapse SN. The lifetimes of these massive stars are computed using the stellar evolution tracks of \citet{2012A&A...537A.146E}; the SN mass and metal yields are computed using the work of \citet{2007PhR...442..269W} while all SNe are assumed to have an explosion energy of $10^{51}$ erg.  This cluster of stars is embedded in an initially static, homogeneous ISM, with each SN occurring at the same location. The resulting superbubble is evolved using a 1D, spherically symmetric, Lagrangian hydrodynamic solver first developed by \citet{2016ApJ...821...76D}. Cells are split (merged) when they become sufficiently larger (smaller) than the average resolution. Metallicity-dependent cooling (assuming collisional ionization equilibrium) is included using \texttt{GRACKLE} \citep{2017MNRAS.466.2217S}. The simulation is evolved until the radial momentum reaches a maximum, at which point it is assumed that the superbubble mixes into the ISM.

\subsection{3D simulations}
\label{section:code:3D}

\begin{table*}
\caption{ Initial Conditions. The mass resolution $\Delta m$ is not included for the 1D run, as it is neither constant in space nor time.
\label{tab:ics}
}
\begin{tabular}{lccrcrrrrrrc}
Name & 1D/3D & $B_{z,0}$ & $\beta$ & $\Delta x_0$ & $\Delta m$ &  $L$ \\
 & & ($\mu$G) &   & (pc) & ($M_\odot$) & (pc)   \\
\hline
\texttt{1D\_06\_HD} & 1D  & 0  & $\infty$   & 0.6 &  &  1200   \\  
\texttt{3D\_07\_HD} & 3D  & 0 &  $\infty$ &  0.7 &  0.01 & 300   \\  
\texttt{3D\_10\_HD} & 3D  & 0 &  $\infty$ &  1.0 &  0.03 & 600   \\  
\texttt{3D\_13\_HD} & 3D  & 0 &  $\infty$ &  1.3 &  0.08 & 400   \\  
\texttt{3D\_20\_HD} & 3D  & 0 &  $\infty$ &  2.0 & 0.26 &  600   \\  
\texttt{3D\_40\_HD} & 3D  & 0 &  $\infty$ &  4.0 & 2.10 &  600   \\  
\texttt{3D\_20\_MHD} & 3D  & 5 & 0.05 &  2.0 & 0.26 & 1200      
\end{tabular}
\end{table*}

Rather than adapt our 1D code to work in 3D, we instead chose to use the \texttt{GIZMO} simulation code \citep{2015MNRAS.450...53H, 2016MNRAS.455...51H}, which includes a Lagrangian hydrodynamic solver with additional support for magnetohydrodynamics (MHD). For all of our runs, we used the Meshless Finite Mass solver on a periodic domain, while ignoring the effects of gravity. We assume the gas follows an ideal equation of state with a constant adiabatic index $\gamma= 5 / 3$, as in our 1D simulation. When including magnetic fields, we used \texttt{GIZMO}'s standard solver for ideal MHD, as detailed in \citet{2016MNRAS.455...51H}.

We modify the standard \texttt{GIZMO} code in two ways.\footnote{Our modifications of \texttt{GIZMO} and our analysis routines can be found at: \url{github.com/egentry/gizmo-clustered-SNe}. } First, we added metallicity-dependent cooling using \texttt{GRACKLE} \citep{2017MNRAS.466.2217S}.  Second, we inject SN ejecta, distributed in time, mass, and metal content using the same realization of SN properties as our 1D simulation. At the time of each SN, we inject new gas particles (each with mass approximately equal to the average existing particle mass) at random locations using a spherical Gaussian kernel with a dispersion of 2 pc centred on the origin.
Each new particle has equal mass and metallicity, which are determined by the SN ejecta yields.\footnote{While this approach leads to a well-sampled injection kernel at our higher resolutions, the kernel is only sampled by about five new particles for each SN in our lowest resolution run, \texttt{3D\_40\_HD}. This undersampling is not ideal and might slightly alter the bubble's evolution, but the stochasticity this introduces does not appear to affect our conclusions.}
For simulations which include magnetic fields, we linearly interpolate the magnetic field vector from nearby existing particles to the origin, and initialize the new feedback particles with that interpolated magnetic field vector. This procedure does not exactly preserve $\nabla\cdot\boldsymbol{B} = 0$, but \texttt{GIZMO}'s divergence cleaning procedure rapidly damps away the non-solenoidal component of the field produced by our injection procedure.

We initialize the 3D simulations with the same static ($\boldsymbol{v}=0$) homogeneous ISM as our 1D simulations ($\rho = 1.33 m_\mathrm{H}$ cm$^{-3}$, $Z=0.02$ and $T \sim 340$ K). For simulations with magnetic fields, we include a homogeneous seed field, with $\boldsymbol{B} = (0, 0, 5)$ $\mu$G (identical to \citealt{2015A&A...576A..95I}), corresponding to a plasma $\beta \approx 0.05$. We place particles of mass $\Delta m$ on an evenly spaced grid of spacing $\Delta x_0$, which extends for a box size of $L$. Particle locations are perturbed on the mpc scale in order to avoid pathologies in \texttt{GIZMO}'s density solver caused by the symmetric grid. In \autoref{tab:ics}, we present the parameters of the initial conditions. We typically\footnote{The only exceptions are simulations \texttt{3D\_07\_HD} and \texttt{3D\_13\_HD}, which cannot be run to completion because they have smaller box sizes in order to minimize computational expense. These simulations are run until the shock approximately reaches the edge of the box. They are not meant to provide final values, but rather to enable us to investigate convergence of the results up to the times when these runs end.} run each 3D simulation for 40 Myr, by which point the radial momentum, the quantity of primary interest for our study, had stabilized. We also carry out a smaller set of simulations in which we give the ISM a larger initial perturbation, whose magnitude shows a proper physical dependence on resolution. We describe these simulations in Appendix \ref{section:perturbations}, where we show that their results are nearly identical to those of our fiducial simulations. For this reason, we will not discuss them further in the main text.

\section{Results}
\label{section:results}

\begin{table*}
\caption{ Results.}
\label{tab:results}
\begin{tabular}{lccrrrrrrrrr}
Name & $N_\mathrm{SNe}$ &  t  & $R_\mathrm{eff}$ & $M_\mathrm{affected}$ & $p_\mathrm{max} / N_\mathrm{SNe}$ & $p_\mathrm{ratchet} / N_\mathrm{SNe}$ &  $p_{t=6.46\text{ Myr}} / N_\mathrm{SNe}$ & $E_\mathrm{kin}$ & $\Delta E_\mathrm{int}$ \\
 & & (Myr) & (pc) & ($ 10^6 M_\odot)$ & ($100 M_\odot$ km s$^{-1}$ ) & ($100 M_\odot$ km s$^{-1}$ ) & ($100 M_\odot$ km s$^{-1}$ ) & ($10^{49}$ erg) & ($10^{49}$ erg) \\
\hline
\texttt{1D\_06\_HD} & 11 & 94.8 & 552 & 23.2 & 33978 & 33978 & 5987 & 65.0 & 26.3 \\
\texttt{3D\_07\_HD} & 11 & -- & -- & -- & -- & -- & 1027 & -- & -- \\
\texttt{3D\_10\_HD} & 11 & 30.7 & 218 & 1.7 & 2425 & 2474 & 948 & 8.7 & 0.8 \\
\texttt{3D\_13\_HD} & 11 & -- & -- & -- & -- & -- & 911 & -- & -- \\
\texttt{3D\_20\_HD} & 11 & 30.7 & 200 & 1.5 & 2128 & 2182 & 862 & 7.4 & 0.8 \\
\texttt{3D\_40\_HD} & 11 & 31.7 & 209 & 1.8 & 2007 & 2039 & 901 & 6.8 & 7.0 \\
\texttt{3D\_20\_MHD} & 11 & 29.6 & 423 & 10.5 & 1213 & 2418 & 1020 & 12.6 & 13.1
\end{tabular}
\end{table*}

In \autoref{tab:results}, we provide a summary of the key numeric results of each simulation.  First, we extract the time of maximum momentum after the last SN (only accurate within about $0.5$ Myr.) At that time we extract the effective radius of the region affected by the bubble (particles with speeds greater than 1 m s$^{-1}$)
\begin{equation}
	R_\mathrm{eff} = \left(\frac{3}{4 \pi} \sum_{i : |\boldsymbol{v}_i| > 1 \text{m s}^{-1} } \mathrm{Volume_i} \right)^{1/3}
\end{equation}
and the total mass of those particles, $M_\mathrm{affected}$.\footnote{
The exact velocity threshold is somewhat arbitrary, leading to roughly 10 per cent uncertainty in the affected mass depending on the chosen threshold.}
Next we extract the kinetic energy $E_\mathrm{kin}$ and the change in the internal energy $\Delta E_\mathrm{int}$ of the entire domain (which should be approximately equal to the values for the bubble-affected region). Finally, we extract the radial momentum using three approaches: one by simply measuring the radial momentum at the same time as the previous quantities (denoted $p_\mathrm{max}$), another using a ``ratchet'' approach justified and explained in \autoref{section:results:MHD} (denoted $p_\mathrm{ratchet}$), and third by extracting the momentum at the last time achieved by all simulations ($t=6.46$ Myr).

In the following subsections we discuss the results in greater detail. First, we compare our 1D and 3D results in \autoref{section:results:3D}. Second, we look at the effect of including magnetic fields in our 3D simulations in \autoref{section:results:MHD}.

\subsection{Hydrodynamic results and convergence study}
\label{section:results:3D}

\begin{figure}
\includegraphics[width=\columnwidth]{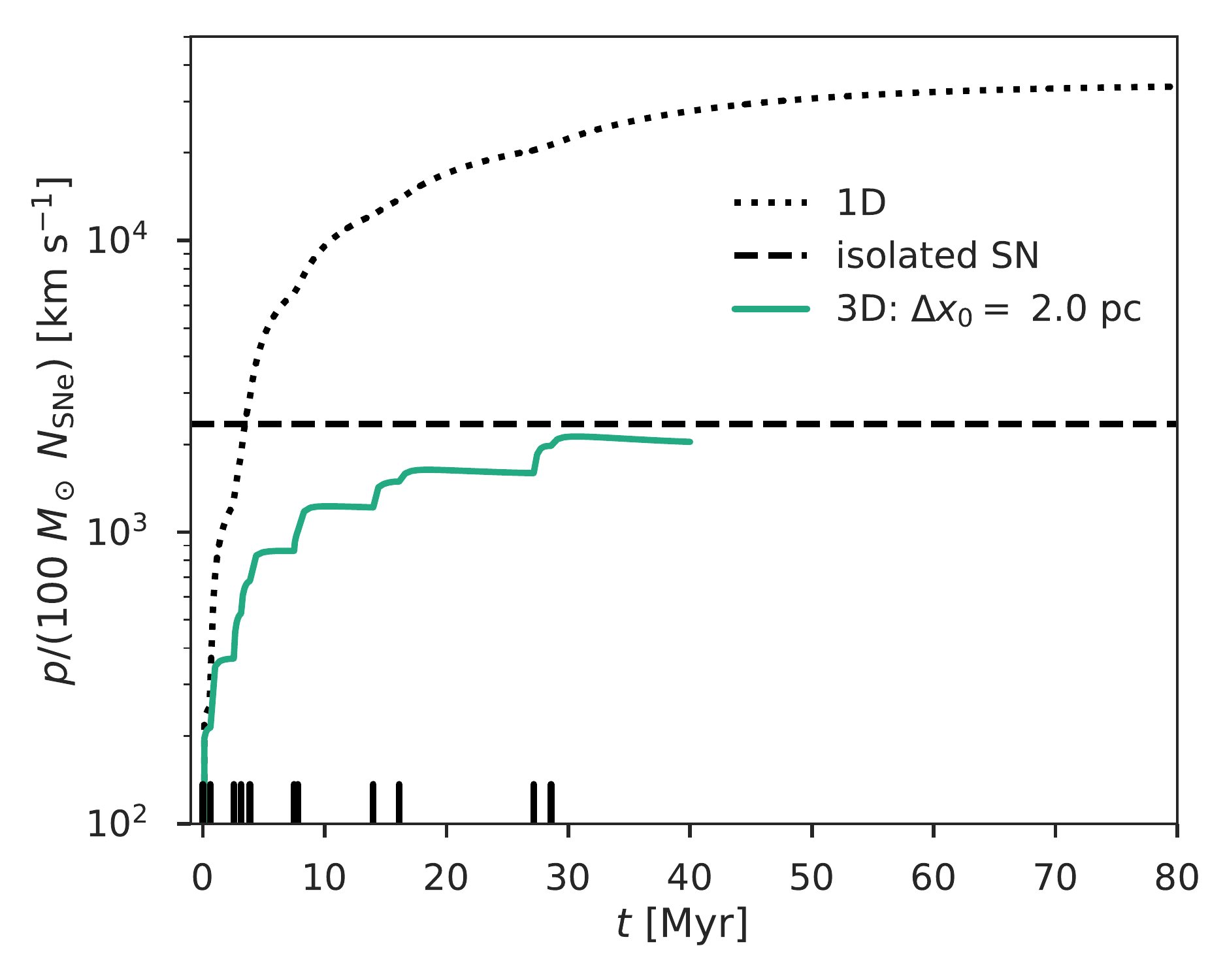}
\caption{
Comparison of the momentum evolution of 1D and 3D simulations of the same cluster (simulations \texttt{1D\_06\_HD}, and \texttt{3D\_20\_HD}, respectively). The `isolated SN' value is estimated using the first SN of the \texttt{3D\_20\_HD} simulation, although it does not vary substantially between any of our 3D simulations.
}
\label{fig:resolution_study:1D_vs_3D}
\end{figure}

\begin{figure}
\includegraphics[width=\columnwidth]{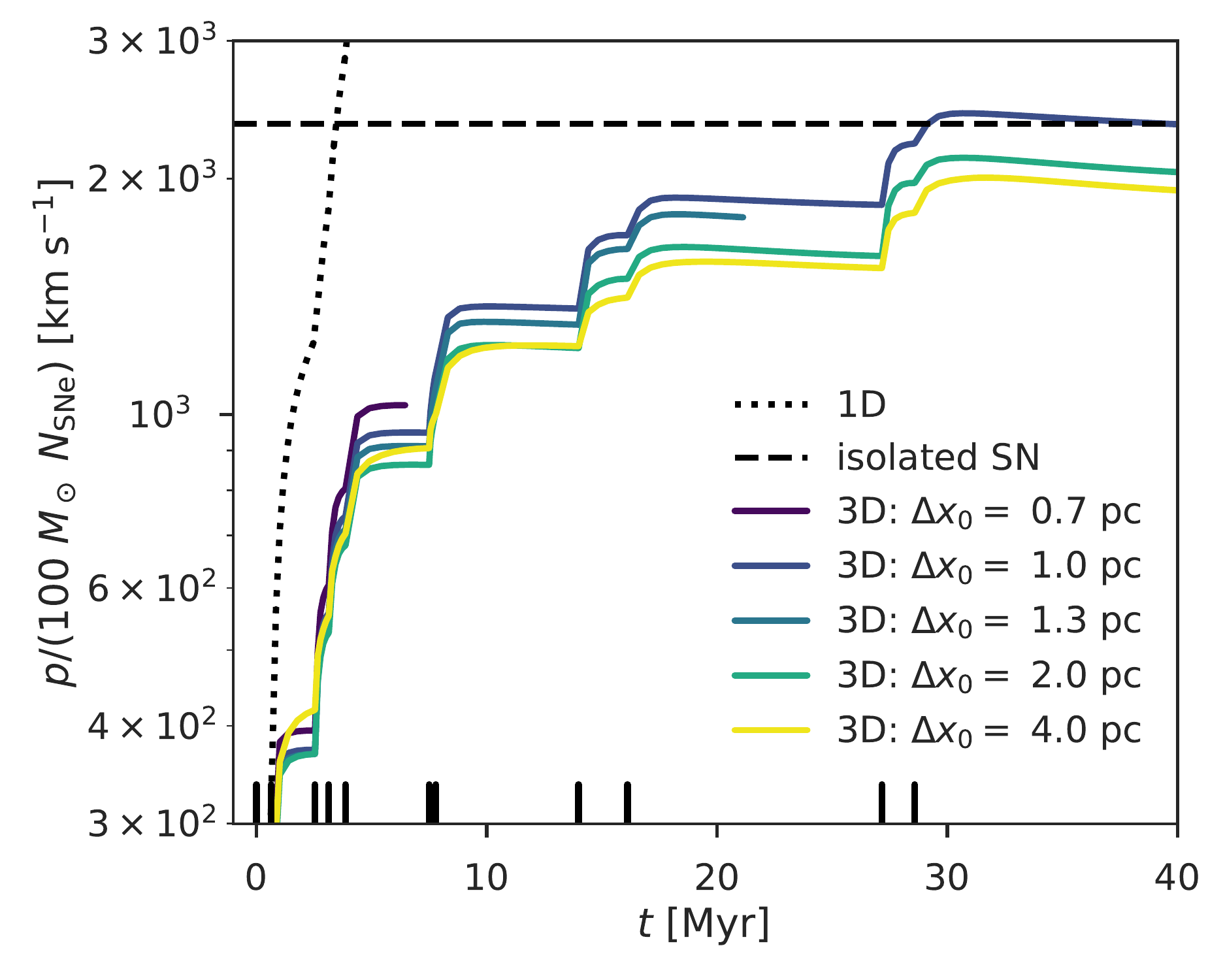}
\caption{
Resolution study of our 3D HD simulations.
}
\label{fig:resolution_study:HD}
\end{figure}

\begin{figure}
\includegraphics[width=\columnwidth]{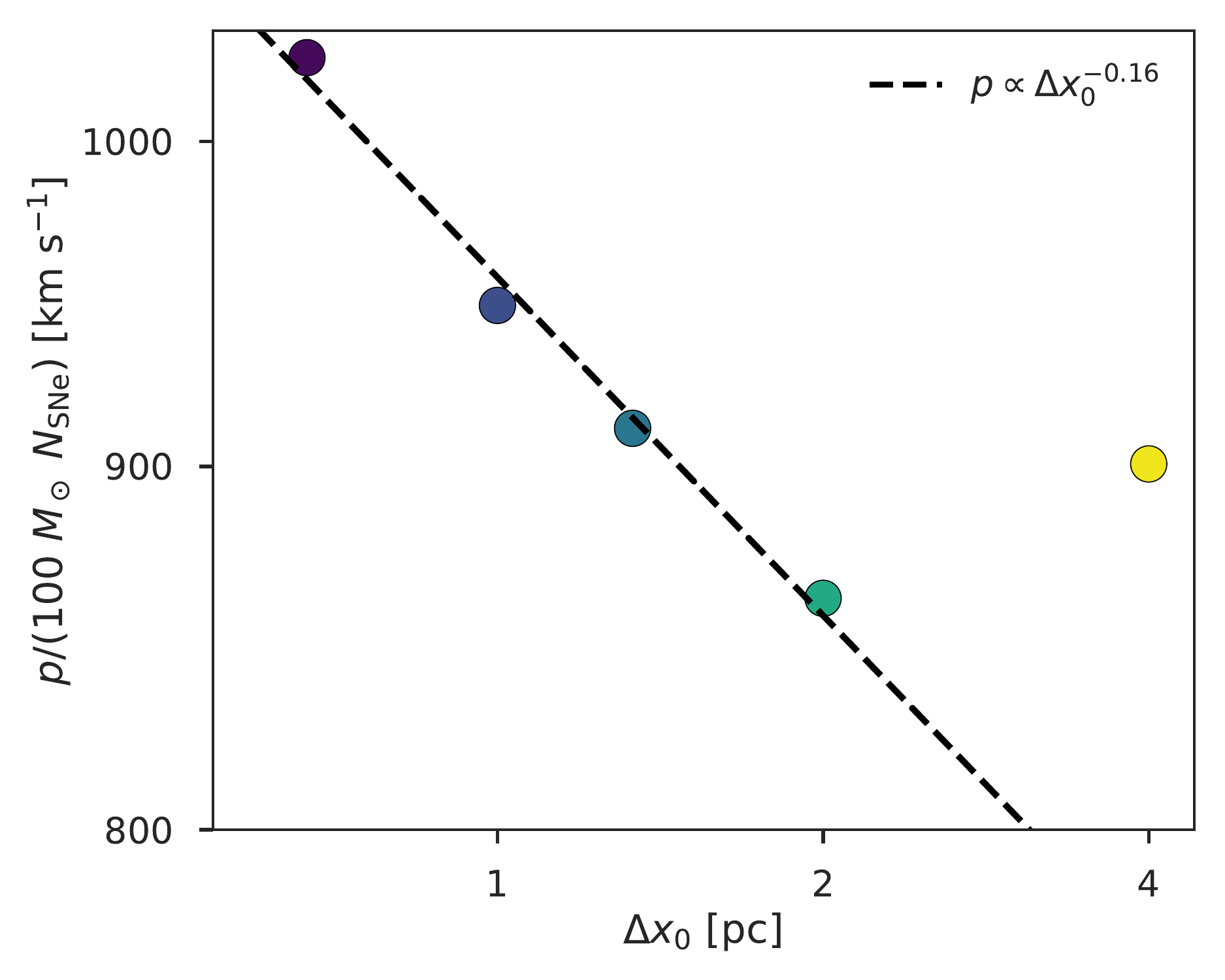}
\caption{Resolution study of our 3D HD simulations at the last time achieved by all simulations. Colours are consistent with the resolution study figures above. The black dashed line shows the best power-law fit to all 3D HD simulations \emph{except} the worst resolution simulation (\texttt{3D\_40\_HD}). Both axes are plotted using log scales. 
}
\label{fig:outlier}
\end{figure}

\begin{figure}
\includegraphics[width=\columnwidth]{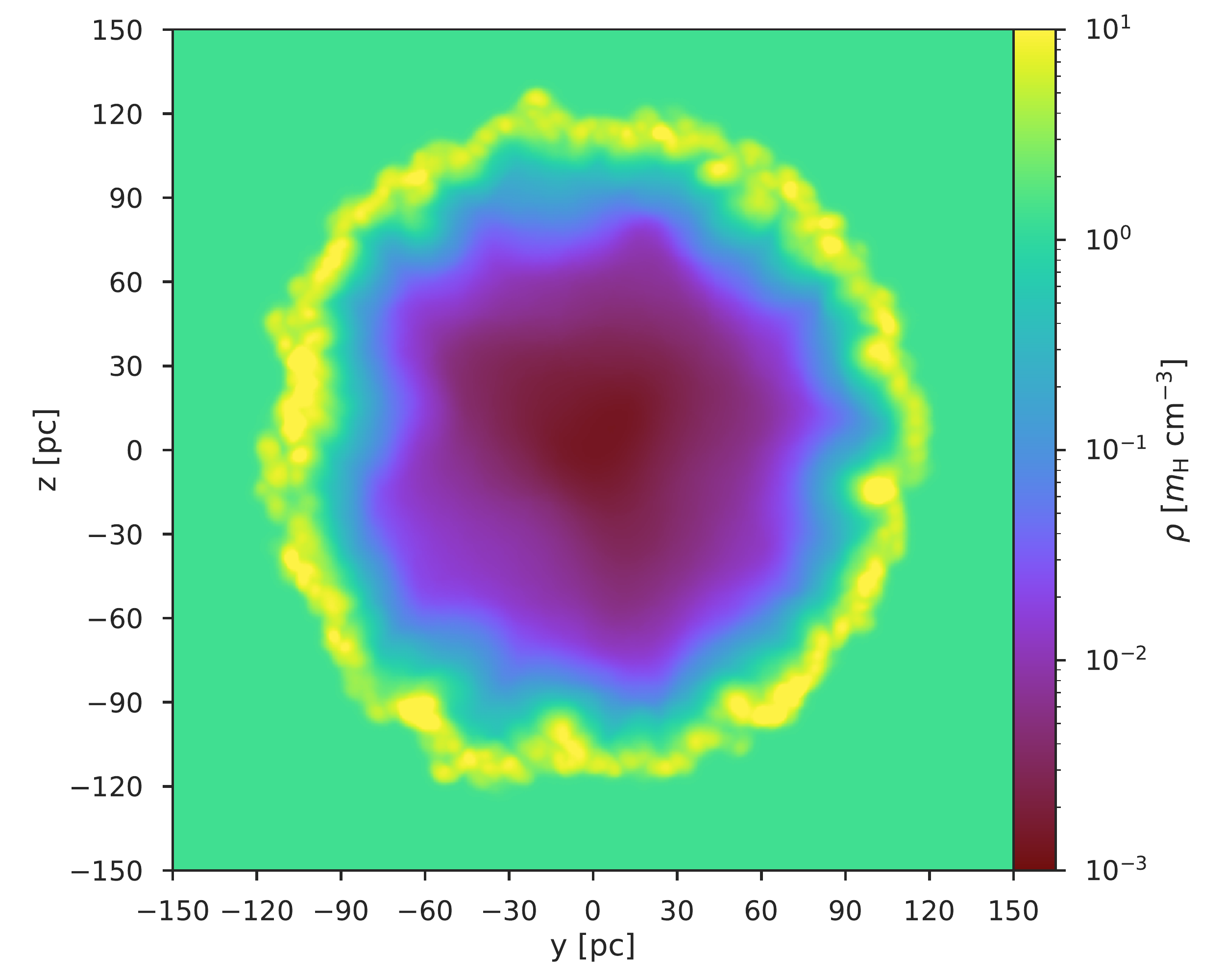}
\caption{
Reference density slice of the median resolution completed 3D simulation (\texttt{3D\_20\_HD}) at $t=7.53$ Myr, approximately $0.03$ Myr after the sixth SN.
}
\label{fig:slice:HD}
\end{figure}

\begin{figure*}
\centering
\includegraphics[width=0.93\textwidth]{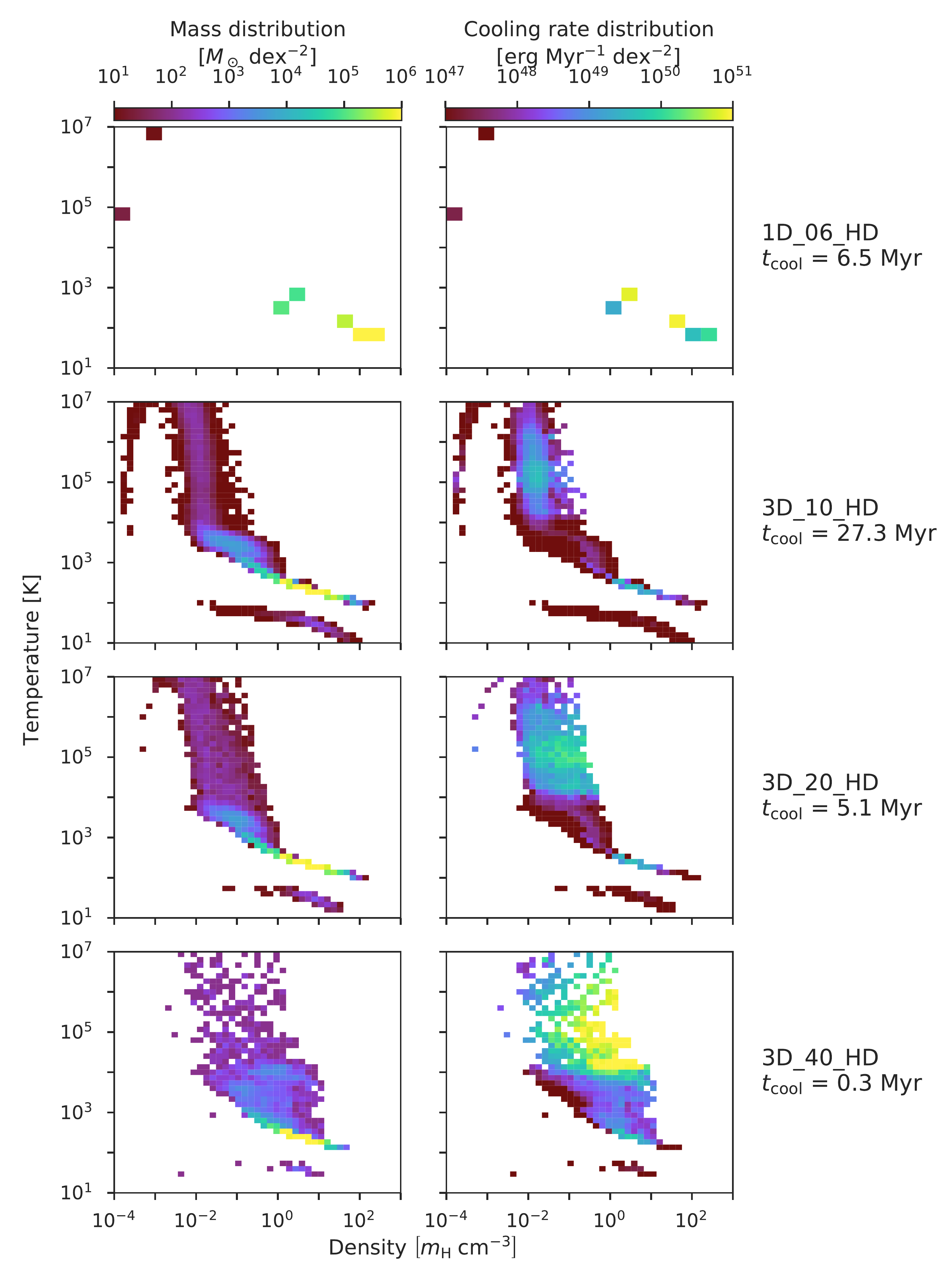}
\caption{
Phase diagrams for our completed HD simulations at $t=7.53$ Myr, about $0.03$ Myr after the sixth SN, when all simulations still retain almost all of the energy from the most recent SN.  The left column shows the distribution of mass within temperature-density space, and the right column shows the cooling rate distribution within the space.  The rows show the non-magnetized simulations with initial resolution worsening from top to bottom. To the right of each row, we give the cooling time of each simulation, $t_\mathrm{cool} \equiv E_\mathrm{int} / \dot{E}_\mathrm{cool}$, for reference.
}
\label{fig:phase:HD}
\end{figure*}

In \autoref{fig:resolution_study:1D_vs_3D}, we show the radial momentum evolution of our median-resolution completed 3D simulation without MHD (\texttt{3D\_20\_HD}), and compare it to our 1D simulation. As can be seen in that figure, we observe a significant difference between the final momenta in our 1D and 3D simulations. While our 1D simulation of clustered SNe shows a large gain in momentum per SN compared to the isolated SN yield,\footnote{
We estimate the isolated SN momentum yield, $2.4 \times 10^5$ $M_\odot$ km s$^{-1}$, using the first SN of our \texttt{3D\_20\_HD} simulation, although all of our 3D simulations would give the same value within a few percent. This is approximately consistent with previous single SN simulations (e.g. \citealt{2015MNRAS.450..504M,2015ApJ...802...99K}).} 
our 3D simulation shows no such gain.  That discrepancy needs to be addressed.

This cannot be explained just by the fact that the 3D simulation has a lower initial resolution. In our previous work we tested the resolution dependence in our 1D simulations, and found that even with an initial spatial resolution of 5 pc, we still measured a terminal momentum yield roughly 10 times higher than what we find in our 3D simulation here as long as we ran our code in pseudo-Lagrangian rather than Eulerian mode \citep[Figure 14]{2017MNRAS.465.2471G}. So the problem is not convergence in our 1D simulation, but we have not yet shown whether our 3D results are converged.

To test for convergence in our 3D simulations, we compare our simulations which differ only in resolution (\texttt{3D\_07\_HD}, \texttt{3D\_10\_HD}, \texttt{3D\_13\_HD}, \texttt{3D\_20\_HD}  and \texttt{3D\_40\_HD}); in \autoref{fig:resolution_study:HD}, we show the momentum evolution of each simulation.  From that figure, we conclude that our 3D simulations do not appear converged, unlike our 1D simulations. The terminal momentum yield is increasing monotonically as we increase the resolution, so our 3D results are converging in the direction of our 1D results, but even at the highest resolution we can afford the momentum yield remains well below the 1D case. Thus we do not know if the 3D results would converge to the same value as the 1D case, even with infinite resolution.

We further illustrate the non-convergence in \autoref{fig:outlier}, which shows the momentum of the shell at 6.46 Myr, the latest time we are able to reach at all resolutions. As the figure shows, with the exception of the lowest resolution run there is a clear trend of increasing momentum at higher resolution; we discuss possible explanations for the anomalous behaviour of the lowest resolution run in Appendix \ref{section:outlier}. A simple power-law fit to the points at resolutions of $\Delta x_0 = 2$ pc or better suggests that the momentum is increasing with resolution as $p \propto \Delta x_0^{-0.16}$. If we naively extrapolate this trend to the peak initial resolution of 0.03 pc achieved in our 1D simulations, the predicted momentum would be a factor of $\sim 2$ larger than the highest resolution run shown, though this may well be an underestimate since \autoref{fig:outlier} shows that the momentum appears to increase with resolution somewhat faster than predicted by a simple power-law fit. In any event, it is clear that, even at a resolution of 0.7 pc, our results are not converged.

To gain additional insight into the resolution-dependence of our results, and the differences between the 1D and 3D runs, we show a density slice through the centre of simulation \texttt{3D\_20\_HD} at $t = 7.53$ Myr shown in \autoref{fig:slice:HD}. Clearly in 3D, the interface between the hot bubble interior and the cold shell is not spherically symmetric. These anisotropies are the result of physical instabilities (such as the Vishniac instabilities and the Rayleigh--Taylor instability) amplifying numerical inhomogeneities in the background ISM \citep{1983ApJ...274..152V,1994ApJ...428..186V,1988ApJ...324..776M,1993ApJ...407..207M,2013A&A...550A..49K,2016MNRAS.456..710F,2017MNRAS.465.1720Y}. To see how this might affect the terminal momentum, we turn to density--temperature phase diagrams which are shown in \autoref{fig:phase:HD}. These phase diagrams correspond to a time soon after the sixth SN, with a delay long enough to allow the injected energy to spread throughout the bubble but sufficiently short to avoid significant energy losses due to cooling in any simulation. 
All 3D simulations have about $1.1 \times 10^{51}$ erg more total energy than the start of the simulations, but \texttt{1D\_06\_HD} retains more energy from previous SNe, and contains about $2.7 \times 10^{51}$ erg of total energy relative to the simulation start. When we look at the mass-weighted phase diagram for our highest resolution completed simulation (\texttt{3D\_10\_HD}), we see that the mass is dominated by a cold dense shell, with a minority of mass in less-dense warm and hot phases ($>10^3$ K).\footnote{We also see a negligible amount of mass at unusually low temperatures, $<100$ K. These particles are SN ejecta, which have very high metallicities that have been frozen-in due to the Lagrangian nature of the code.}  Even when we vary the resolution, we only find negligible changes in the fraction of mass in the cold phase; the cold phase ($T<10^3$ K) contributes 99.2\% of the affected mass in every completed simulation of our resolution study (\texttt{3D\_10\_HD}, \texttt{3D\_20\_HD}, and \texttt{3D\_40\_HD}).  What does change is the density and temperature distribution of the warm/hot gas ($T>10^3$ K). As we increase the resolution, the warm/hot gas shifts to lower and lower densities. This effect is very apparent for gas near the peak of the cooling curve (specifically $3 \times 10^4$ K $ < T < 3 \times 10^{5}$ K), which has a mass-weighted median density of $\sim 10^{-1}$ $m_\mathrm{H}$ cm$^{-3}$ in our lowest resolution run, and $\sim 10^{-2}$ $m_\mathrm{H}$ cm$^{-3}$ in our highest resolution completed run.

This has a significant impact on the overall cooling times of the simulations. The right column of \autoref{fig:phase:HD} shows that while the cold, dense phase dominates the mass, the minority of mass in the warm/hot phases dominates the cooling rate. This is important because resolution primarily affects these warm/hot phases, and it affects those phases by shifting them to higher densities at lower resolution, causing each particle to become more efficient at cooling.  This results in significantly shorter cooling times: from 27 Myr at the best resolution to 0.3 Myr at the worst resolution, nearly two orders of magnitude difference. This increase primarily occurs in the warm/hot phases; at all resolutions gas warmer than $10^3$ K constitutes slightly less than 1\% of the total mass, but this mass is responsible for 81\% of the cooling at our highest resolution completed run, and $>99\%$ of the cooling at our lowest resolution.

When we look at the phase diagrams for our 1D simulation, we see significant differences in the distributions of mass and cooling rate, leading to the very different behaviour of the 1D simulation. In particular, the 1D simulation completely lacks material at intermediate densities ($\sim 10^{-2} - 10^0$ $m_\mathrm{H}$ cm$^{-3}$) due to how well the 1D simulation retains the contact discontinuity. The diffuse bubble-dense shell transition occurs within only a few cells, and the entire dense shell is resolved by just 5-10 cells.  In our 3D simulations, these intermediate densities contribute a negligible amount of mass, but are responsible for much of the cooling. Without this intermediate phase material, almost all of the cooling in the 1D simulation occurs in the dense shell. We defer further discussion about the physical nature of the intermediate-temperature gas, and to what extent its properties are determined by physics versus numerics in the various simulations, in \autoref{section:analysis}.

\subsection{Magnetic fields}
\label{section:results:MHD}

\begin{figure}
\includegraphics[width=\columnwidth]{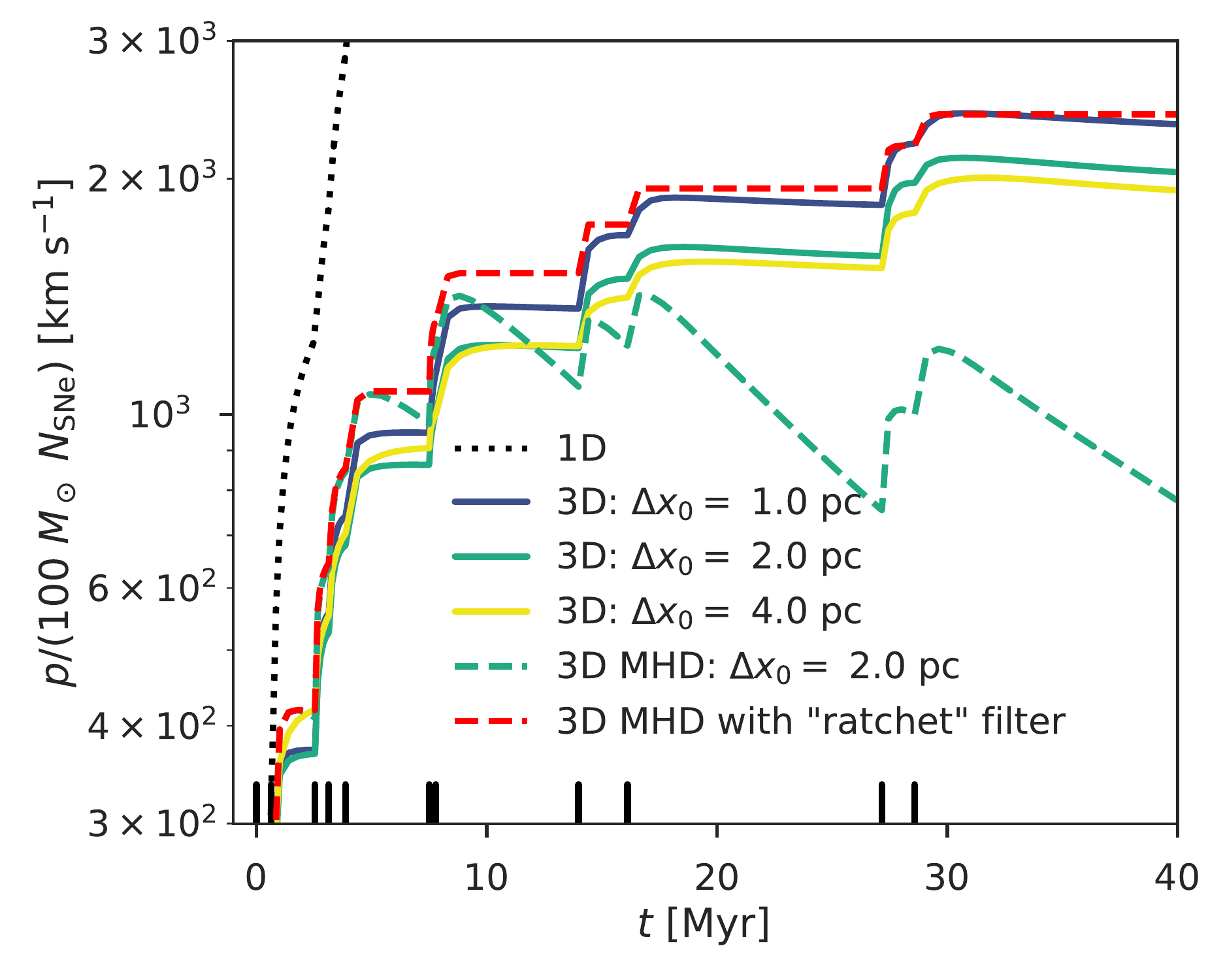}
\caption{
Same as \autoref{fig:resolution_study:HD}, except now including the momentum evolution of our MHD simulation (\texttt{3D\_20\_MHD}; blue dashed curve), as well its ``ratchet''-filtered momentum evolution, $p_\mathrm{ratchet}$ (red dashed curve), and excluding our incomplete HD runs (\texttt{3D\_07\_HD} and \texttt{3D\_13\_HD}). 
}
\label{fig:resolution_study:MHD}
\end{figure}

\begin{figure}
\includegraphics[width=\columnwidth]{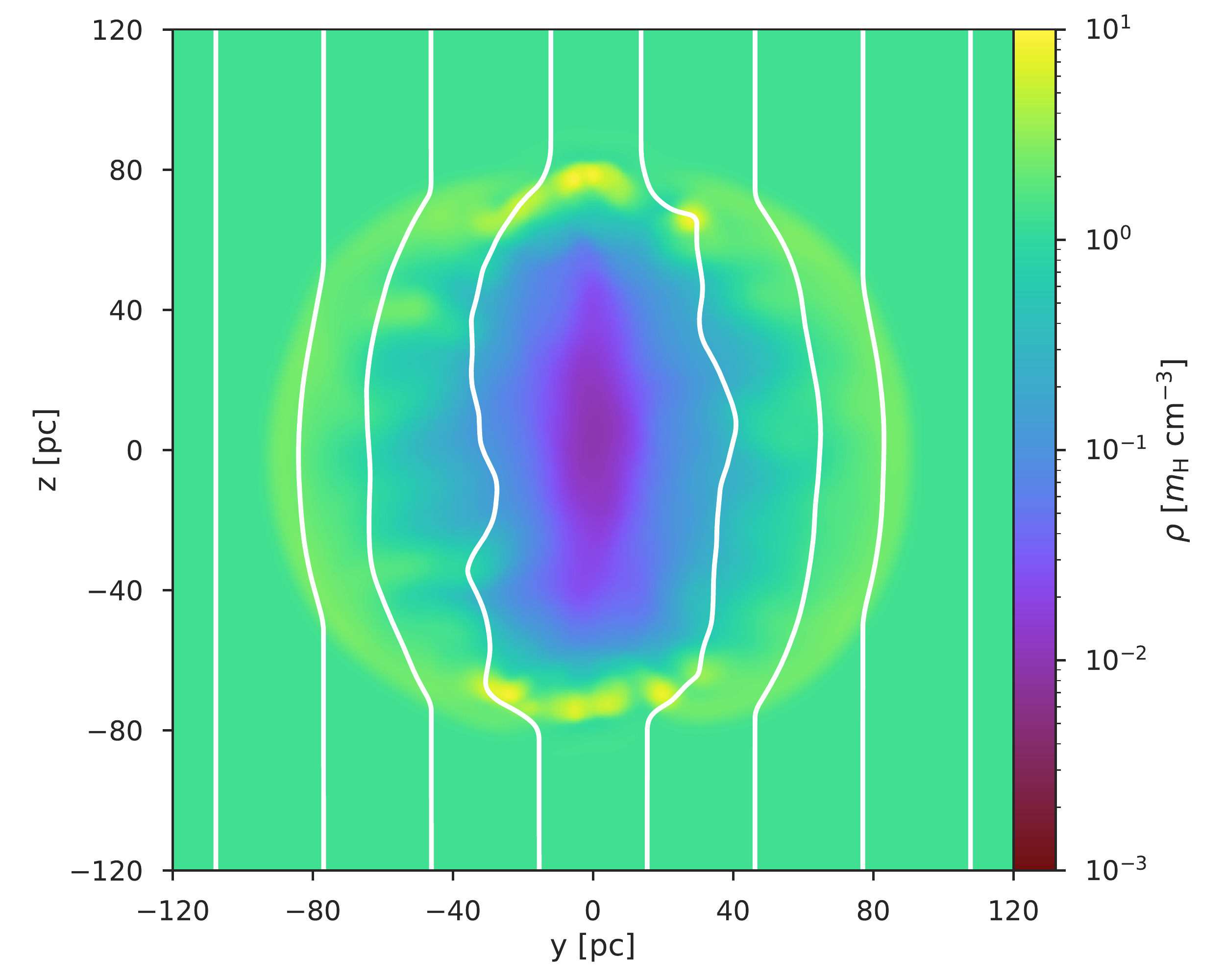}
\caption{
Same as \autoref{fig:slice:HD}, except now for simulation \texttt{3D\_20\_MHD} with \emph{approximate} magnetic field lines overplotted, and at an earlier time ($t = $ 2.56 Myr; approximately 0.02 Myr after the third SN).
}
\label{fig:slice:MHD}
\end{figure}

\begin{figure*}
\centering
\includegraphics[width=0.93\textwidth]{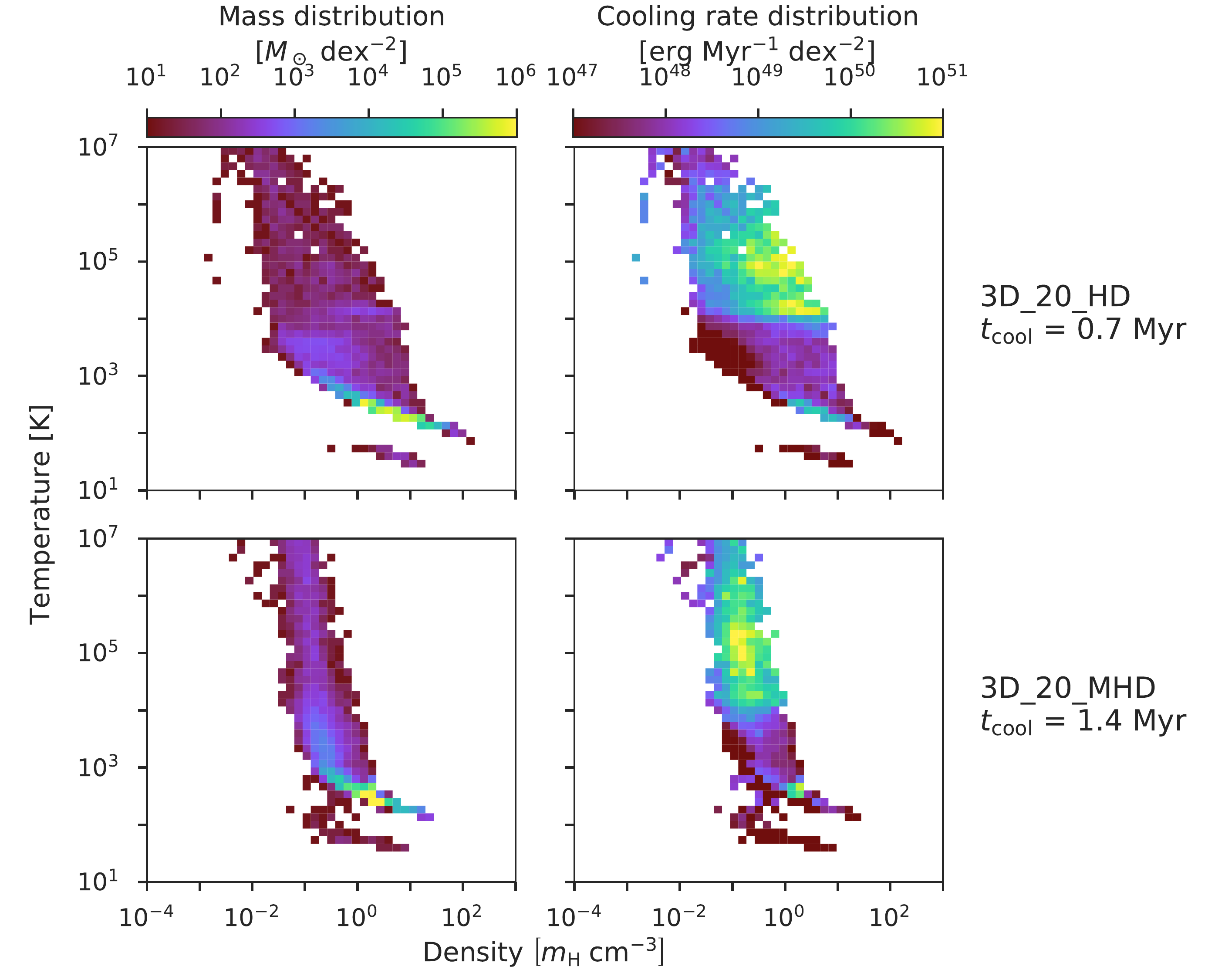}
\caption{
Same as \autoref{fig:phase:HD}, except at the earlier time shown in \autoref{fig:slice:MHD} ($t=2.56$ Myr), and only showing the magnetized simulation (\texttt{3D\_20\_MHD}) and the corresponding non-magnetized simulation with the same resolution (\texttt{3D\_20\_HD}).
}
\label{fig:phase:MHD}
\end{figure*}

In \autoref{section:results:3D}, we showed that our numerical methods and resolution are not sufficient to achieve converged values of final radial momentum and other key parameters due to physical instabilities that develop within the superbubble shell.  As described in \autoref{section:intro}, we expect that magnetic fields might affect the growth of physical instabilities, so we also run an MHD simulation as described in \autoref{section:code:3D} to test the impact of magnetic fields on the final momentum.  While the more standard method of extracting momentum, $p_\mathrm{max}$, quoted in \autoref{tab:results} appears to show that the inclusion of magnetic fields significantly decreases the final momentum, in this subsection we show that that method for estimating the asymptotic momentum (finding the maximum momentum following the last SN) is an oversimplification for simulations with magnetic fields. When we better isolate the momentum added by SNe, we find that adding magnetic fields can actually increase the momentum yield at fixed resolution. Indeed, our $\Delta x_0 = 2.0$ pc MHD run produces a larger momentum injection than our $\Delta x_0 = 1.0$ pc HD run.

First, to illustrate why the interpretation of the MHD simulation is more complex, in \autoref{fig:resolution_study:MHD} we compare its momentum evolution to those of the non-magnetized simulations.  The MHD simulation initially shows an \emph{increased} momentum yield relative to the corresponding simulation without magnetic fields at the same resolution (\texttt{3D\_20\_MHD}), but then the momentum decreases due to magnetic tension forces. The reason for this is obvious if we examine a density slice at an earlier time, \footnote{
We chose to look at an earlier snapshot, when the magnetization has only perturbed the bubble structure, rather than the later time shown in \autoref{fig:slice:HD}, when the magnetization would have caused a strong, non-linear change in the structure which could not be treated as a perturbation. In both cases the magnetic tension is present, but the earlier time makes it more straightforward to compare to the non-magnetized runs.}
as shown in \autoref{fig:slice:MHD}: the expanding shell bends magnetic field lines outward, and the field lines exert a corresponding magnetic tension that reduces the radial momentum of the expanding shell.
This effect is so strong that the momentum peaks after just seven SNe; the remaining four SNe clearly add momentum but not enough to overcome the steady decline.

Due to this effect, the quantity $p_{\rm max}$ (the maximum momentum after the last SN) that we have used to characterize the hydrodynamic simulations is somewhat misleading, since our goal is to study the momentum injected by SNe, not the combined effects of SNe and magnetic confinement. To avoid this, we define an alternative quantity $p_{\rm ratchet}$. To compute this quantity we sum any positive changes in radial momentum between snapshots, while ignoring any negative changes. We plot $p_{\rm ratchet}$ in \autoref{fig:resolution_study:MHD}, and report the final value in \autoref{tab:results}. As expected, for the non-magnetic runs $p_{\rm ratchet}$ and $p_{\rm max}$ are essentially the same, and thus examining $p_{\rm ratchet}$ allows us to make an apples-to-apples comparison between the magnetic and non-magnetic results.

This comparison is revealing, in that it shows that our simulation with magnetic fields (\texttt{3D\_20\_MHD}) injects \emph{about 10\% more} momentum than the analogous simulations without magnetic fields (\texttt{3D\_20\_HD}). 
The full explanation for this difference will likely be complicated -- for example, the bubble morphology and phase structure are significantly altered at late times relative to the non-magnetized runs -- but we can see if our results are at least consistent with the hypothesis that magnetic fields could inhibit the growth of instabilities, leading to less phase mixing and cooling.  To test this hypothesis, we compare phase diagrams for the resolution-matched magnetized and non-magnetized runs in \autoref{fig:phase:MHD}, shown at the same time ($t=2.56$ Myr) as \autoref{fig:slice:MHD}.  There we see that magnetic fields have an effect similar to that of increasing resolution in \autoref{fig:phase:HD}: both suppress the growth of fluid instabilities, causing the material near the peak of the cooling curve to stay at lower densities where it cools less efficiently. For gas near the peak of the cooling curve (specifically $3 \times 10^4$ K $ < T < 3 \times 10^{5}$ K), the median density of the non-magnetized run is $1.7 \times 10^{-1}$ $m_\mathrm{H}$ cm$^{-3}$, while in the magnetized run it is $1.4 \times 10^{-1}$ $m_\mathrm{H}$ cm$^{-3}$ -- a modest change, but a change in the predicted direction. As a result the overall cooling time is about two times longer in the MHD run. Thus by suppressing the growth of instabilities, the inclusion of magnetic fields results in a longer overall cooling time which should contribute to a higher yield of momentum.

\section{Analysis}
\label{section:analysis}

In \autoref{section:results}, we showed our broad results, which have three key features:
(1) Our 1D Lagrangian simulation finishes with about 10 times more momentum than our 3D simulations, and is converged with respect to resolution.
(2) Our 3D HD simulations show a general increase in momentum as resolution improves, but are not converged even at the highest resolutions we can reach [similar to the 1D Eulerian simulations of \citet{2017MNRAS.465.2471G}, which are not converged even at a resolution of 0.31 pc].
(3) Our MHD simulations show less momentum than the resolution-matched HD simulation when the momentum is estimated directly, but more momentum when our ``ratchet'' filter is used.

The phase diagrams shown in \autoref{fig:phase:HD} and \autoref{fig:phase:MHD} reveal that the changes in momentum budget appear to be associated with changes in the total mass and mean density of gas at temperatures of $\approx 10^5$ K, near the peak of the cooling curve, which dominates the cooling budget. In this section we seek to understand the physical origin of these differences, with the goal of understanding whether the converged 1D or non-converged 3D results are likely to be closer to reality.

\subsection{What determines convergence or non-convergence?}

\begin{figure}
\includegraphics[width=\columnwidth]{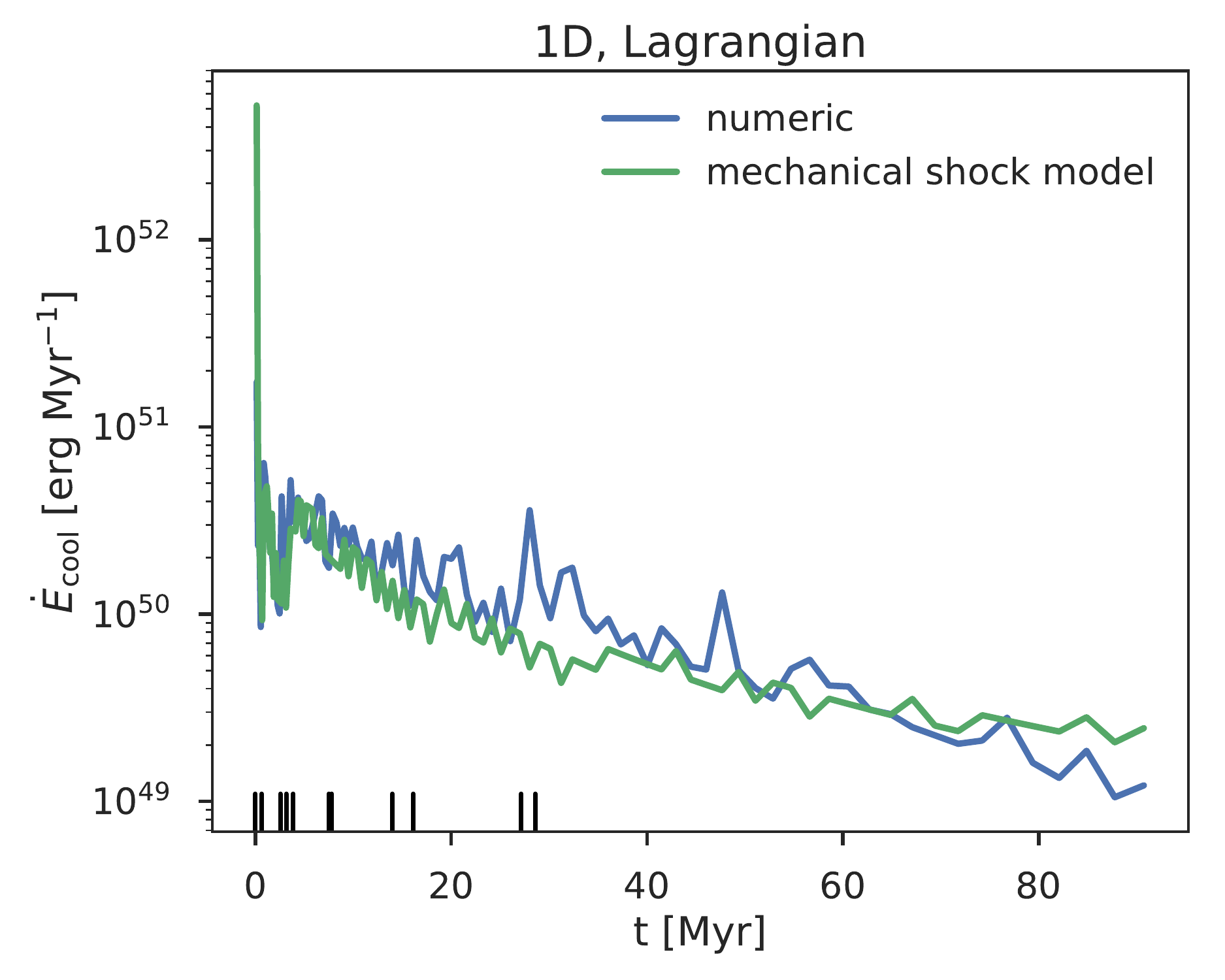}
\includegraphics[width=\columnwidth]{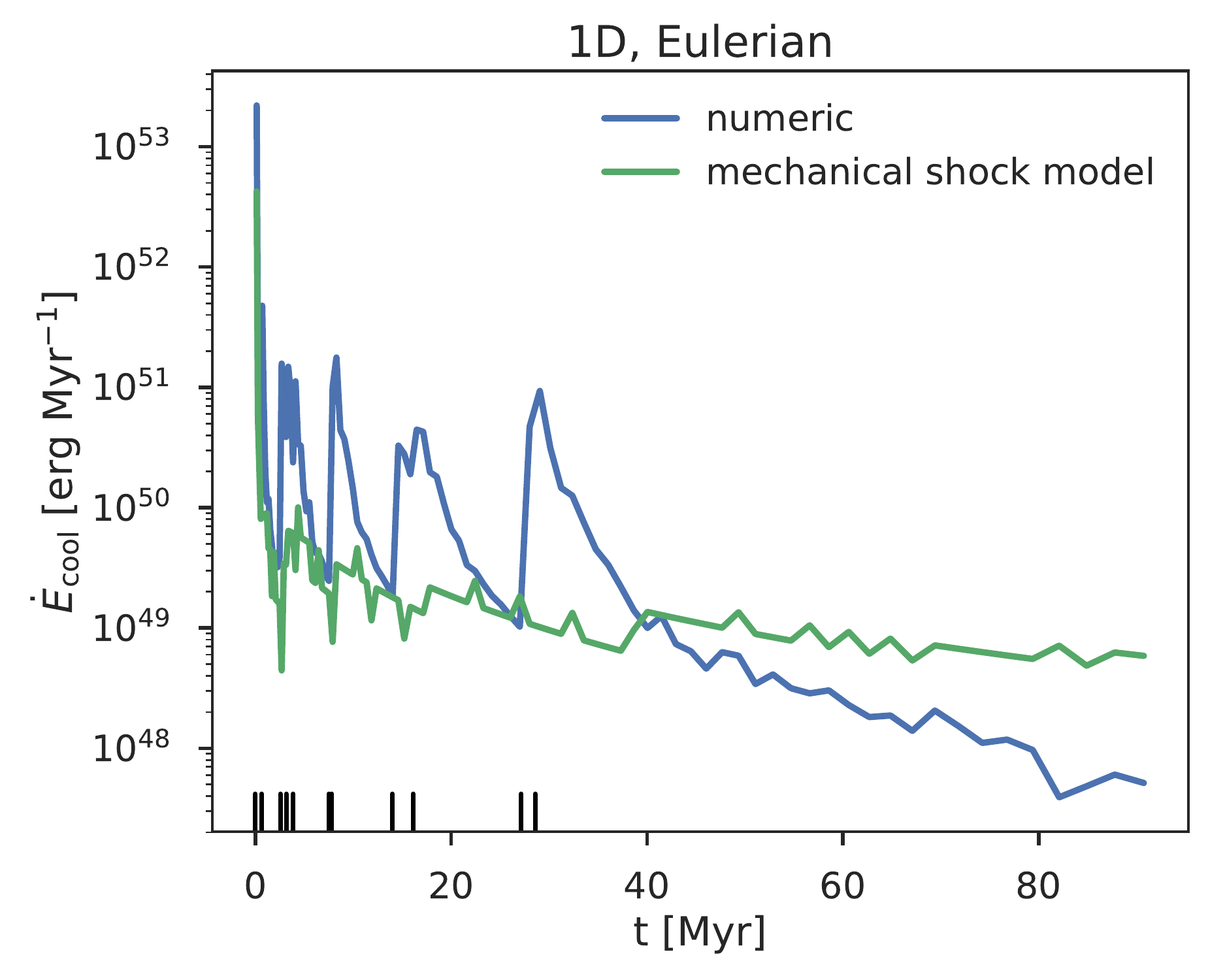}
\includegraphics[width=\columnwidth]{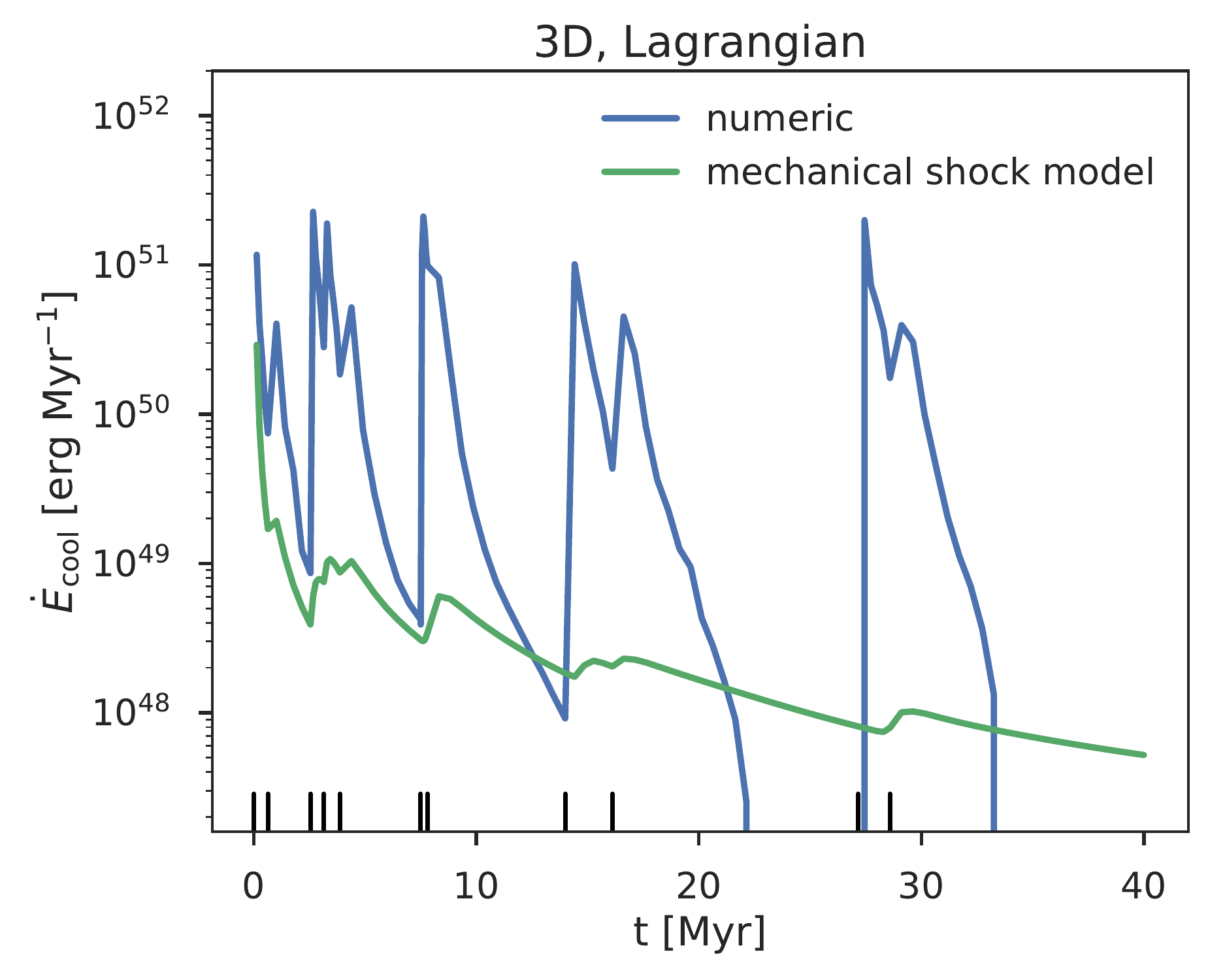}
\caption{Comparison of the numeric cooling rate with the cooling rate predicted by our mechanical shock model, \autoref{eq:shock_cooling}, for our 1D Lagrangian simulation (\texttt{1D\_06\_HD}; \textit{top}), our 1D Eulerian simulation (\texttt{1D\_06\_HD}, but run with the code in Eulerian mode; \textit{middle}), and our 3D HD simulation with 2 pc initial resolution (\texttt{3D\_20\_HD}; \textit{bottom}).
}
\label{fig:shock}
\end{figure}

As a first step in this analysis, we investigate why our 1D Lagrangian simulations are converged while our 3D simulations are not. A simplistic view of superbubble cooling is one where the diffuse bubble interior contains most of the thermal energy but is radiatively inefficient, while the cold dense shell \emph{is} radiatively efficient but does not have significant amounts of thermal energy to radiate. The cooling rate is then set by how quickly energy can transfer from one phase to the other.

The minimum amount of energy transfer comes from the fact that the hot overpressured bubble is doing work on the shell, transferring thermal energy from the interior into kinetic energy of the shell. As the shell sweeps up and shocks new material, some of this kinetic energy will be transferred into thermal energy within the shell, where it can be easily radiated.  To lowest order, we predict this mechanical process would result in the following cooling rate:

\begin{equation}
	\dot{E}_\mathrm{cool, mechanical}	= 4 \pi R_\mathrm{shock}^2 V_\mathrm{shock} \rho_0 \left(\frac{V_\mathrm{shock}^2}{2} \right)
	\label{eq:shock_cooling}.
\end{equation}

This expression assumes a supersonic shock, and that all of the energy that is converted from kinetic to internal energy is immediately radiated away.  At each simulation snapshot, we can compute $R_\mathrm{shock}$ and $V_\mathrm{shock}$
\footnote{
For our 3D simulations, we estimate $R_\mathrm{shock}$ as the mean radius of the overdense particles and $V_\mathrm{shock}$ as the mean radial velocity of the overdense particles. For our 1D simulations, we determine $R_\mathrm{shock}$ as the outermost overdense cell \citep[see][]{2017MNRAS.465.2471G}, and determine $V_\mathrm{shock}$ by taking the difference of $R_\mathrm{shock}$ between snapshots.
} and compute the expected cooling rate using \autoref{eq:shock_cooling}. We can then compare that to the observed cooling rate, calculated by \texttt{GRACKLE} for each snapshot.

We begin our predicted-versus-actual cooling rate comparisons with our 1D Lagrangian simulation (\texttt{1D\_06\_HD}) shown in the top panel of \autoref{fig:shock}.  In that figure we can see that even though our mechanical shock model is simple, it does a generally good job predicting the observed cooling rate. On the other hand, we can repeat this with a simulation that is identical to \texttt{1D\_06\_HD} except it uses an \emph{Eulerian} hydrodynamic solver, leading to the results shown in the middle panel of \autoref{fig:shock}. This reveals a very different picture: there are many times when the observed cooling rate is over an order of magnitude greater than our mechanical shock model would predict. And when the observed rate is lower than predicted, it is because the shell has already transitioned from a non-linear shock to a linear sound wave, for which we know \autoref{eq:shock_cooling} should not hold.  While the mechanical shock model can explain most of the behaviour behind the 1D Lagrangian simulation, in the 1D \emph{Eulerian} simulation the chosen numerical methods lead to much higher cooling rates which must be powered by additional thermal energy being pumped into the shell. When we apply this same approach to one of our 3D Lagrangian simulations (specifically simulation \texttt{3D\_20\_HD}, shown in the bottom panel of \autoref{fig:shock}), we find a behaviour similar to the 1D Eulerian simulation and very different from the 1D Lagrangian simulation: the actual cooling rates often far exceed the rate predicted by our mechanical shock cooling model.

This analysis makes it clear why the 1D Lagrangian simulations are converged: the radiative cooling rate has reached the minimum allowed by the physical situation of an adiabatic fluid doing work on a medium with a short radiative cooling time. Consequently, increasing the resolution cannot further reduce the rate of radiative loss; it is already as low as physically allowed. If we run the same problem in 1D Eulerian mode, or in 3D but at much lower resolution, the cooling rate is far in excess of the minimum. Cooling is powered not primarily by adiabatic compression of the cold gas followed by radiative loss, but by direct transfer of thermal energy between the hot and cold phases without doing any mechanical work. The rate of transfer is clearly resolution-dependent, which explains why the 1D Eulerian and 3D simulations are not converged.

\subsection{Conduction and numerical mixing across the interface}

Since the key difference between the converged 1D Lagrangian simulations and the unconverged 3D simulations is the relative importance of energy transfer by mechanical work versus other mechanisms, we next investigate the expected rate of non-mechanical energy transfer in reality, and how that compares to the rate in our simulations.

In a bistable radiative medium such as the one we are simulating, conductive transfer occurs across an interface whose characteristic width, known as the Field length, is given by \citep{1990ApJ...358..375B}
\begin{equation}
\lambda_F = \left(\frac{\kappa T}{n^2 \Lambda}\right)^{1/2},
\end{equation}
where $\kappa$ is the thermal conductivity and $\Lambda$ is the cooling function. The conductive heat flux is $F\sim \kappa T/\lambda_F$, so the total rate at which energy conducts across an interface of area $A$ and is lost to radiation is
\begin{equation}
\label{eq:e_dot_cond}
\dot{E}_{\rm cond} \sim \dot{E}_{\rm cool} \sim A \frac{\kappa T}{\lambda_F}.
\end{equation}
\autoref{fig:phase:HD} shows that, for simulation \texttt{3D\_40\_HD} at time $t=7.53$ Myr, typical values for the gas that dominates the cooling are $n=1$ cm$^{-3}$ and $T=4 \times 10^{5}$ K. Using \citeauthor{1990ApJ...358..375B}'s expression for thermal conductivity, assuming no suppression by magnetic fields and no saturation, together with the approximate cooling function $\Lambda$ from \citet[their equation~4, which we use for simplicity, rather than performing the full \texttt{GRACKLE} calculation]{2002ApJ...564L..97K}, we find $\lambda_F \approx 0.003$ pc. Using the lower density $n\approx 10^{-1}$ cm$^{-3}$ found in our highest resolution completed 3D simulation (\texttt{3D\_10\_HD}) would increase this to $\lambda_F \approx 0.03$ pc. By contrast, our best 3D simulation resolution is an order of magnitude larger; only our 1D Lagrangian simulation approaches this resolution. Thus the true physical width of the interface is far from resolved in any of our 3D simulations.

\begin{figure}
\includegraphics[width=\columnwidth]{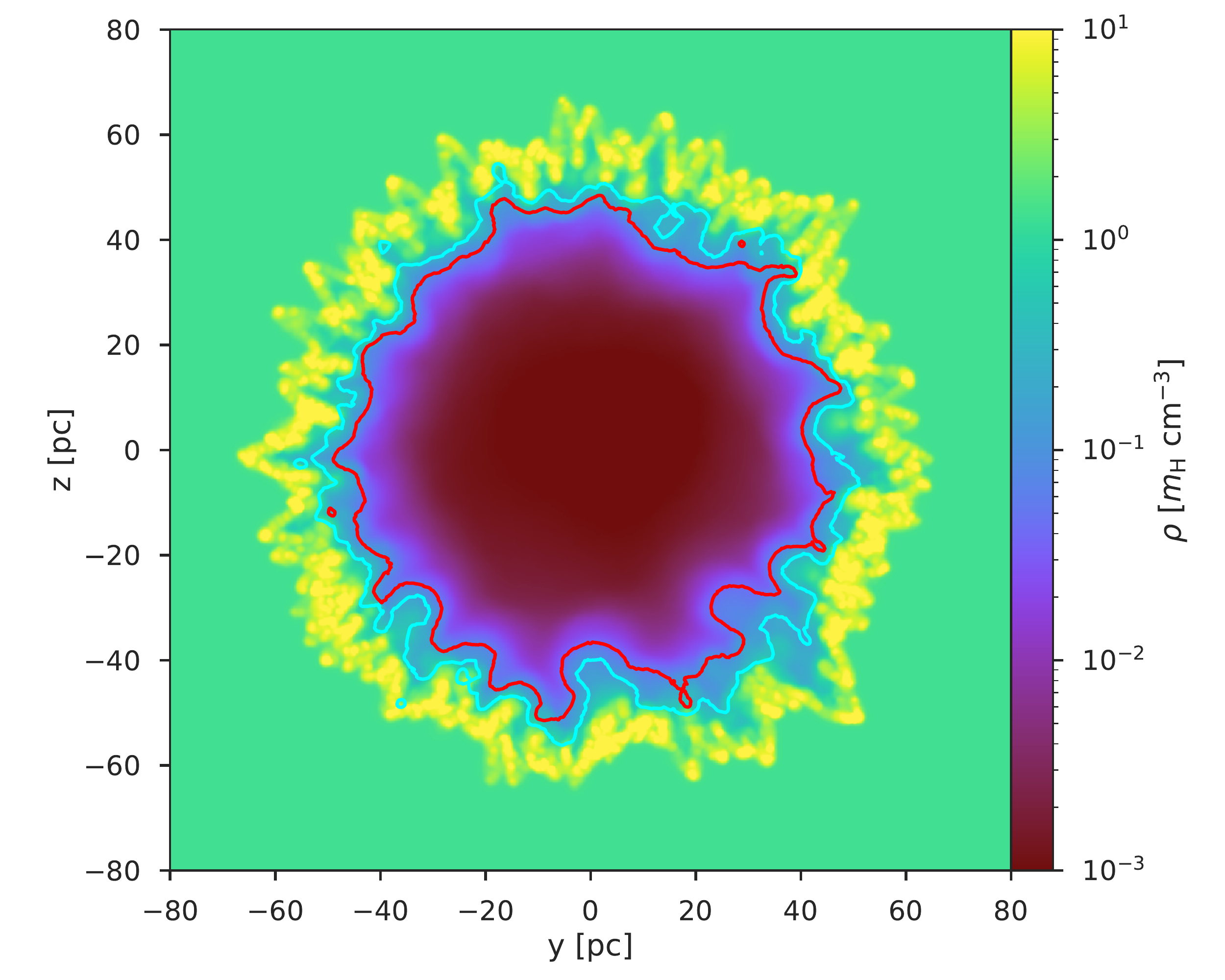}
\includegraphics[width=\columnwidth]{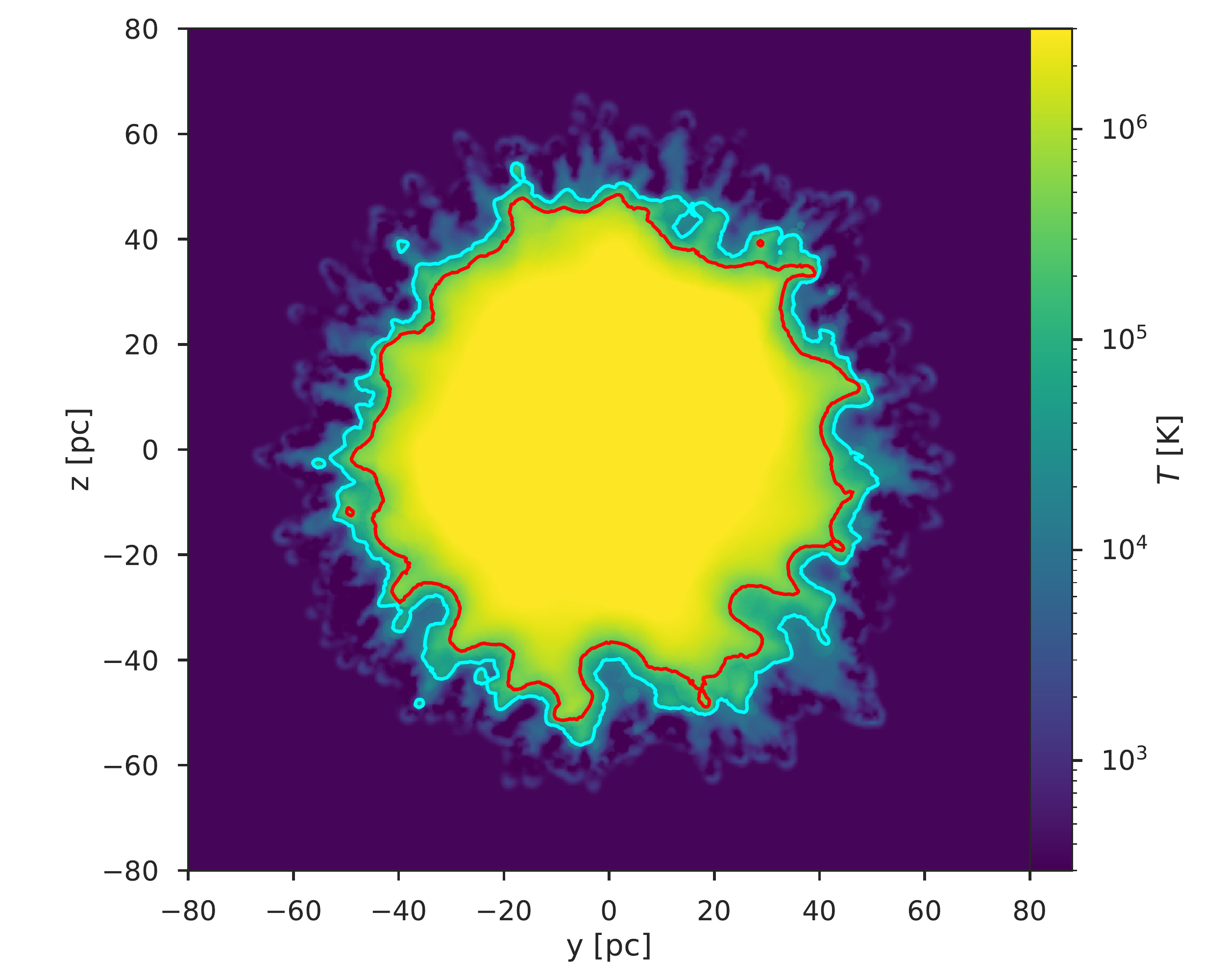}
\caption{
Density (top) and temperature (bottom) slices of the highest resolution 3D simulation (\texttt{3D\_07\_HD}) at $t=1.01$ Myr, approximately $0.5$ Myr after the second SN. The lighter cyan and darker red contours correspond to temperatures $3\times 10^4$ K and $3 \times 10^5$ K, respectively, which are roughly the bounds of the peak of the cooling curve.}
\label{fig:slice:temperature}
\end{figure}

\begin{table*}
\caption{
Interface properties at $t=1.01$ Myr ($0.5$ Myr after the second SN). Here $N_{\rm paticles}$ and $m$ are the number of particles and total mass in the interface (defined as the set of particles with temperatures in the range $3\times 10^4 - 3\times 10^5$ K, near the peak of the cooling curve), $r_{\rm median}$ is the median radius of interface particles, $h_\mathrm{median}$ is the median scale length of interface particles, $\Delta r_\mathrm{IQR}$ is the interquartile range of interface particle radii.
\label{tab:interface}
}
\begin{tabular}{lcccccc}
Name & $N_{\rm particles}$ & $m$ & $r_\mathrm{median}$ & $h_{\rm median}$ & $\Delta r_\mathrm{IQR}$ & $\Delta r_\mathrm{IQR}/h_\mathrm{median}$ \\
& & ($M_\odot$) & (pc) & (pc) & (pc) \\
\hline
\texttt{3D\_40\_HD} & 16    & 33.6 & 37.0 & 15.4 & 8.5 & 0.6 \\
\texttt{3D\_20\_HD} & 159  & 41.8 & 39.1 & 9.8 & 8.3 & 0.8 \\
\texttt{3D\_13\_HD} & 485  & 37.8 & 43.3 & 6.6 & 6.4 & 1.0 \\
\texttt{3D\_10\_HD} & 1303 & 42.8 & 45.4 & 4.9 & 5.1 & 1.0 \\
\texttt{3D\_07\_HD} & 4436 & 43.2 & 47.4 & 3.3 & 4.3 & 1.3 \\
\end{tabular}
\end{table*}

In our simulations, as opposed to reality, the width of the interface is set by numerical resolution. We illustrate this point in \autoref{fig:slice:temperature}, which shows temperature and density slices from our highest resolution simulation (\texttt{3D\_07\_HD}) shortly after the second SN. We summarize the physical properties of the hot--cold interface, defined as material between $3 \times 10^4$ K and $3 \times 10^5$ K, roughly corresponding to the peak of the cooling curve, in \autoref{tab:interface}; we include \texttt{3D\_40\_HD} for completeness, but warn that, at this early time, its interface is poorly sampled by only 16 particles, and thus the results for it are not particularly meaningful. The main conclusion to make from \autoref{fig:slice:temperature} and \autoref{tab:interface} is that the physical width of the interface region is of order a particle smoothing length, so the width of the interface is determined by numerics rather than physics.

What is the impact of this underresolution on the rate of radiative loss? Though we do not include explicit conduction (nor would it matter if we did, since our failure to resolve $\lambda_F$ would lead us to greatly underestimate the true conduction rate), any finite-resolution numerical method necessarily has some conduction-like dissipation at the resolution scale. It is convenient to characterize this dissipation in terms of the effective P\'eclet number of the method, which is related to the effective thermal conductivity of the numerical scheme $\kappa_{\rm num}$ by
\begin{equation}
\mathrm{Pe} \sim \frac{Lv n k_B}{\kappa_{\rm num}},
\end{equation}
where $L$ and $v$ are the characteristic length and velocity scales. The exact value of the P\'eclet number will depend on the numerical method. In Eulerian methods where effective conduction is due to fluids mixing at the resolution scale $\Delta x$, we expect to have $\mathrm{Pe} \sim 1$ for $L \sim \Delta x$. In a Lagrangian method $\mathrm{Pe}$ will be substantially larger, since mixing is suppressed. Ignoring this complication, if we replace $\kappa$ with $\kappa_{\rm num}$ and $\lambda_F$ with the interface width $\lambda_I$ in \autoref{eq:e_dot_cond}, the numerical conductive transport and cooling rates are then
\begin{equation}
\dot{E}_{\rm cond, num} \sim \mathrm{Pe}^{-1} A n k_B T v \left(\Delta x/\lambda_I\right).
\end{equation}
Using the same values of $n$ and $T$ given above, the approximate velocity $v \approx 40$ km s$^{-1}$ for the shock at the time shown in \autoref{fig:slice:temperature}, and our empirical finding that $\Delta x/\lambda_I\sim 1$, we find that for a method with $\mathrm{Pe} = 1$ at the resolution scale, $\dot{E}_{\rm cond,num} \approx 10 \dot{E}_{\rm cond}$, i.e., underresolving the interface causes us to overestimate the rate of energy loss by a factor of $\sim 10$. This overcooling problem is substantially reduced for the 1D Lagrangian simulation, since compared to the other simulations it has both a smaller value of $\Delta x/\lambda_I$ (due to its high resolution) and a larger value of $\mathrm{Pe}$ (due to its Lagrangian method). Conversely, this analysis strongly suggests that the ultimate reason for non-convergence in our 3D simulations is that their rates of energy loss are dominated by resolution-dependent artificial conduction. Thus the terminal momentum in these simulations must be regarded as a lower limit on the true value.

\subsection{The role of 3D instabilities}

Before accepting the conclusion that artificial conduction is the culprit for our non-convergence, however, we should examine an alternative hypothesis. The total conductive transport rate (\autoref{eq:e_dot_cond}) depends not just on the conductive flux, but on the area of the interface. Examining \autoref{fig:slice:temperature}, it is clear that the area of the interface is affected by 3D instabilities that are not properly captured in the 1D simulations. Could the non-convergence in 3D be a result of the area not being converged, rather than the conductive flux not being converged? This hypothesis might at first seem plausible, because many instabilities (such as the Rayleigh--Taylor, Richtmyer--Meshkov and Vishniac) initially grow fastest at the smallest scales \citep[e.g.,][]{Taylor192,doi:10.1002/cpa.3160130207,1983ApJ...274..152V,2012ApJ...759...78M}. If the area of the interface is determined by the amount of time that it takes perturbations to grow from the resolution scale that might explain why our highest resolution simulation has the lowest cooling rate: because it had the smallest perturbations to start, and it has the smallest interface area later on, and thus the smallest rate of conduction.

However, we can ultimately rule out this hypothesis for two reasons. First, if the rate of mixing and radiative loss were set by processes developing from grid-scale perturbations, then changing the initial perturbation strength and scaling should have a noticeable impact on the cooling rate. However, as shown in Appendix~\ref{section:perturbations}, our non-convergence is quite robust to the details of the grid-scale perturbations. The results are not any more converged when we impose perturbations whose power spectral density is independent of resolution over all resolved scales, and increasing the initial perturbation strength by a factor of $>25$ has negligible effects on the outcome. Second, once they are strongly non-linear, interface instabilities are typically dominated by larger rather than smaller modes. Examining \autoref{fig:slice:temperature}, it is clear that even just after the second supernova we already have strongly non-linear perturbations in the shell, with each spike well resolved by many particles. If linear growth of instabilities from the grid scale were the source of our non-convergence, we would expect to see the greatest resolution dependence at early times, when the perturbations are smallest, and convergence between the runs at later time, when the instabilities reach non-linear saturation. Examining \autoref{fig:resolution_study:HD}, however, shows exactly the opposite pattern: resolution matters more at later times than at earlier ones.

However, simply because we can rule out the hypothesis that the non-convergence of the 3D simulations is a result of our failure to capture the growth of 3D instabilities, it does not follow that the instabilities are not important. \autoref{fig:slice:temperature} clearly shows that the area of the interface in 3D is clearly larger than $4\pi R_{\rm shock}^2$, and thus the rate of conduction across the interface should be higher than it is in our 1D simulations. Thus while our 3D simulations represent a lower limit on the terminal momentum, we must regard the 1D simulations as representing an upper limit, since the interface in 1D has the smallest possible area.


\section{Conclusions}
\label{section:conclusions}

In this paper, we revisit the question of whether clustering of SNe leads to significant differences in the amount of momentum and kinetic energy that supernova remnants deliver to the ISM. This question is strongly debated in the literature, with published results offering a menu of answers that range from a relatively modest increase or decrease \citep{2015ApJ...802...99K,2015MNRAS.451.2757W,2017ApJ...834...25K} to a substantial increase \citep{2014MNRAS.442.3013K,2017MNRAS.465.2471G}. We investigate whether this discrepancy in results is due to numerical or physical effects, and to what extent it might depend on whether the flow is modelled as magnetized or non-magnetized.

Our results offer some encouragement and also some unhappy news regarding the prospects for treating supernova feedback in galactic and cosmological simulations. The encouraging aspect of our findings is that we have identified the likely cause of the discrepancy between the published results. We find that the key physical mechanism driving the differences between our runs, and almost certainly between other published results, is the rate of mixing across the contact discontinuity between the hot interior of a superbubble and the cool gas in the shell around it. Our 1D Lagrangian results \citep{2017MNRAS.465.2471G} maintain the contact discontinuity nearly perfectly, and give it the smallest possible area, and this explains why they produce large gains in terminal momentum per supernova due to clustering. However, these results likely represent an upper limit on the momentum gain, because they do not properly capture instabilities that increase the area of the contact discontinuity and thus encourage mixing across it.

In 3D, both physical instabilities and numerical mixing produce intermediate temperature gas that radiates rapidly and saps the energy of the superbubble, lowering the terminal momentum. Due to this mixing, we are unable to obtain a converged result for the terminal momentum; we find that the terminal momentum continues to increase with resolution even at the highest resolution that we complete (1 pc initial linear resolution, 0.03 $M_\odot$ mass resolution).
The cause of this effect is clear: as we increase the resolution, we find that the mean density and total mass of gas near the peak of the cooling curve continuously decreases (indicating a decrease in mixing), and this typically leads to a decrease in the amount of energy lost to radiation. Consequently, we are forced to conclude that even at our highest resolution in 3D, the mixing and energy transfer rate across the contact discontinuity is dominated by numerical mixing. As a result, our estimate of the momentum per supernova is only a lower limit. 

Our tests with magnetic fields reinforce this conclusion. We find that magnetic fields suppress the growth rate of physical instabilities. This leads an magnetized simulation to inject more momentum per supernova than a non-magnetized simulation, but both still inject far less than the no-mixing case. This is consistent with the conclusion that physical mixing is present in our simulations but numerical mixing is the dominant source. In the real ISM, magnetic fields are doubtless present, so this effect should not be neglected, especially in simulations that are not dominated by numerical mixing.

Our findings cloud the prospects for obtaining a good first-principles estimate of the true supernova momentum yield in a homogeneous ISM. Our peak spatial resolution is higher than that achieved in previous 3D simulations, and we used Lagrangian methods rather than Eulerian methods. We note our choice of Lagrangian rather than Eulerian methods was based on a 1D rather than 3D experiment, and that our results are likely affected by multiple definitions of resolution, such as the mass resolution of ejecta, and not just the peak spatial resolution.
None the less, we are unable to reach convergence. 
We are forced to conclude that the true momentum yield from clustered SNe in a homogeneous ISM remains substantially uncertain. At this point we can only bound it between $\approx 2.4\times 10^5$ $M_\odot$ km s$^{-1}$ per SN (our non-converged 3D result) and $\approx 3.4\times 10^6$ $M_\odot$ km s$^{-1}$ per SN (our converged but 1D result). The 1D result certainly produces too much momentum, since 3D instabilities must enhance the conduction rate at least somewhat by increasing the area of the hot–cold interface. Similarly, our 3D results produce too little momentum, since our 3D results remain dominated by numerical conduction even at the highest attainable resolution; we do not know how close a converged 3D result would lie to the 1D, no-mixing limit.

We conclude by noting that we have not thus far investigated the effects of using a realistically turbulent, multiphase ISM. The presence of density inhomogeneities could well lead to higher rates of mixing across the contact discontinuity, and thus a reduction in the supernova momentum yield. However, we urge caution in interpreting the results of any investigations of these phenomena, since we have shown that even state-of-the-art simulation methods operating at the highest affordable resolutions cannot reach convergence in what should be substantially simpler problems. It is conceivable that the more complex density field of a realistic ISM might make it easier to reach convergence, but such a hope would need to be demonstrated rigorously using convergence studies in multiple numerical methods.

\section*{Acknowledgements}

We thank Peter Mitchell for useful feedback on the role of shock cooling within our 1D simulations. This work was supported by the National Science Foundation (NSF) through grants AST-1405962 (ESG and MRK), AST-1229745 (PM), and DGE-1339067 (ESG), by the Australian Research Council through grant ARC DP160100695 (MRK) and by NASA through grant NNX12AF87G (PM). This research was undertaken with assistance of resources and services from the National Computational Infrastructure (NCI), which is supported by the Australian Government. MRK thanks the Simons Foundation, which contributed to this work through its support for the Simons Symposium `Galactic Superwinds: Beyond Phenomenology'. PM thanks the Pr\'{e}fecture of the Ile-de-France Region for the award of a Blaise Pascal International Research Chair, managed by the Fondation de l'Ecole Normale Sup\'{e}rieure. AL acknowledges support from the European Research Council (Project No. 614199 `BLACK'; Project No. 740120 `INTERSTELLAR').


\bibliographystyle{mnras}
\bibliography{3D_MHD}


\appendix

\section{Sensitivity to Initial Perturbations}
\label{section:perturbations}

In the fiducial 3D simulations presented in the main text, we set up the initial \texttt{GIZMO} particle positions by placing them in a uniform grid and then randomly perturbing each particle position using a Gaussian kernel with a dispersion of $10^{-3}$ times the initial spatial resolution.  This results in an uncorrelated artificial density perturbation with a standard deviation of about $2 \times 10^{-4}$ times the mean density as inferred by \texttt{GIZMO}'s density solver, regardless of resolution.

In order to understand the effect of this perturbation, and how our results depend on its magnitude and whether that magnitude scales with resolution, we rerun a subset of our simulations with an additional perturbation.  In addition to the artificial coordinate-based perturbation, we apply a ``physical-like'' perturbation field directly to the particle masses and densities. To realize this, we generate a white (uncorrelated) Gaussian perturbation field with a magnitude of 5\% of the mean density sampled on the grid of our highest resolution completed simulation (\texttt{3D\_10\_HD}).  For our lower resolution runs, we average the perturbation over appropriately larger apertures, matching what should happen if this were a physical perturbation.  This averaging results in a decreasing perturbation magnitude at worsening resolution, but the magnitude is always at least a factor of 25 larger than the standard coordinate-based perturbation, and the power spectral density of the perturbation is the same at all resolved scales in all simulations. This resolution-dependence is a key difference from the artificial perturbation in our primary runs which has a magnitude that does not change with resolution. 
This process also introduces minor spatial correlations, as some higher resolution particles are equidistant between lower resolution particles, and their perturbation must be shared between multiple lower resolution particles.

\begin{figure}
\includegraphics[width=\columnwidth]{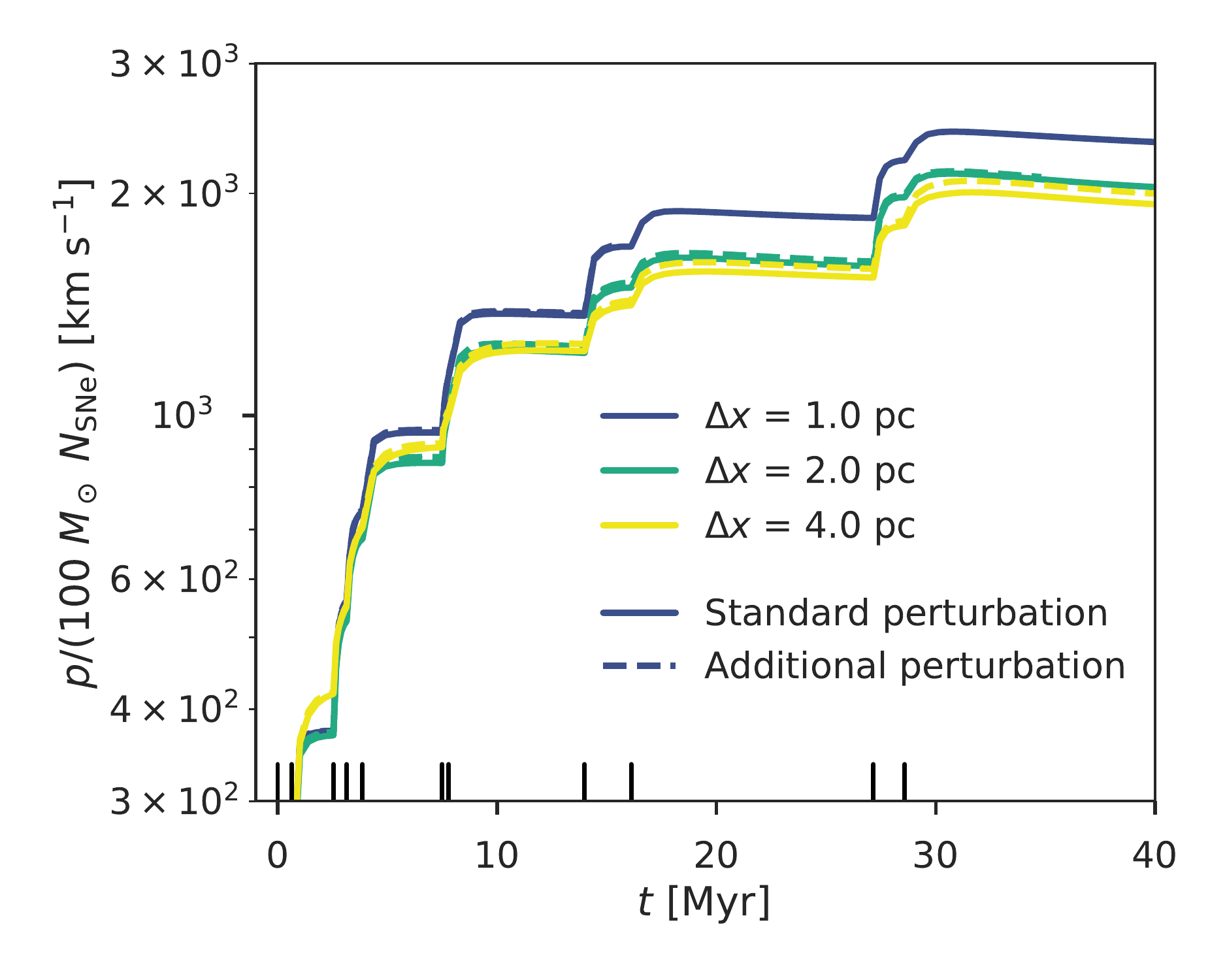}
\caption{Comparison of the momentum evolution of our completed 3D simulations (\texttt{3D\_10\_HD}, \texttt{3D\_20\_HD}, \texttt{3D\_40\_HD}), and similar simulations with an additional, stronger perturbation with magnitudes that correctly scale with resolution. 
}
\label{fig:perturbation}
\end{figure}

In \autoref{fig:perturbation}, we show the results of rerunning our three completed 3D HD simulations (\texttt{3D\_10\_HD}, \texttt{3D\_20\_HD}, and \texttt{3D\_40\_HD}) with these alternative initial conditions.\footnote{The variant of \texttt{3D\_10\_HD} with the additional perturbation has only been run for about 15 Myr due to its computational cost, but we do not expect our conclusions would change if it were run to completion.}  We find that the details of the initial perturbation has very little effect compared to changing the resolution; increasing the perturbation magnitude by a factor of more than 25 has a smaller effect than increasing the spatial resolution by a factor of 2.

\section{Simulation \texttt{3D\_40\_HD} as an outlier at early times}
\label{section:outlier}

In the resolution study (e.g. \autoref{fig:resolution_study:HD}) we see that the momenta of our simulations are well-ordered with respect to resolution at late times but that between the second and third SNe our lowest resolution simulation (\texttt{3D\_40\_HD}) has more momentum than our highest resolution simulation (\texttt{3D\_07\_HD}).
We conjecture that this anomalous behaviour of \texttt{3D\_40\_HD} is related to our SN injection method. As noted in \autoref{section:results:3D}, a typical SN is added using only $\sim 5$ new particles in \texttt{3D\_40\_HD}, leading to an undersampled injection kernel. While it is not clear precisely why undersampling would lead to a systematic increase in momentum, it is strongly suggestive that our simulations start behaving differently right as we hit the resolution limit of one of our methods.

Fortunately, this does not appear to affect our late-time results or our major conclusions. At early times, we recommend treating \texttt{3D\_40\_HD} as an outlier, in which case the momentum will be monotonic with respect to resolution at effectively all times.


\bsp	
\label{lastpage}
\end{document}